\documentclass[twocolumn,preprintnumbers,amsmath,amssymb]{revtex4}

\usepackage[pdftex]{graphicx}
\usepackage{bm}
\usepackage{color}
\usepackage{amsmath}
\usepackage{amsfonts}
\usepackage{amssymb}
\usepackage{epstopdf}
\usepackage{dcolumn}
\usepackage{here}
\usepackage{ulem}

\begin{document}

\preprint{APS/123-QED}

\title{Contrasting Magnetic Structures in SrLaCuSbO$_{6}$ and SrLaCuNbO$_{6}$: Spin-1/2 Quasi-Square-Lattice $\bm {J_1{-}J_2}$ Heisenberg Antiferromagnets}


\author{Masari Watanabe}
\email{watanabe.m.bm@m.titech.ac.jp}
\affiliation{Department of Physics, Tokyo Institute of Technology, Meguro-ku, Tokyo 152-8551, Japan}
\author{Nobuyuki Kurita} 
\affiliation{Department of Physics, Tokyo Institute of Technology, Meguro-ku, Tokyo 152-8551, Japan}
\author{Wataru Ueno}
\author{Kazuki Matsui}
\author{Takayuki Goto}
\affiliation{Physics Division, Sophia University, Chiyoda-ku, Tokyo 102-8544, Japan}
\author{Masato Hagihala}
\affiliation{Materials Sciences Research Center, Japan Atomic Energy Agency, Tokai, Ibaraki 319-1195, Japan}
\author{Hidekazu Tanaka}
\email{tanaka@lee.phys.titech.ac.jp}
\affiliation{Department of Physics, Tokyo Institute of Technology, Meguro-ku, Tokyo 152-8551, Japan}


\date{\today}

\begin{abstract}

We report the magnetic properties of the double perovskites SrLaCuSbO$_6$ (SLCSO) and SrLaCuNbO$_6$ (SLCNO). The temperature dependence of the magnetic susceptibilities of both compounds shows a broad maximum characteristic of an $S\,=\,1/2$ square lattice Heisenberg antiferromagnet. Magnetic ordering occurs at $T_{\rm N}\,{=}\,13.6$ and 15.7 K for SLCSO and SLCNO, respectively. Neutron powder diffraction measurements reveal contrasting spin structures in both compounds. The spin structures of SLCSO and SLCNO below $T_{\rm N}$ are N\'{e}el antiferromagnetic and collinear antiferromagnetic, respectively. This result demonstrates that the nearest-neighbor interaction is dominant in SLCSO, whereas the next-nearest-neighbor interaction is dominant in SLCNO. The magnitude of the ordered moment was evaluated at 3.5 K to be $m\,{=}\,0.39(3)\,\mu_{\rm B}$ for SLCSO and $0.37(1)\,\mu_{\rm B}$ for SLCNO, which are significantly smaller than those calculated using linear spin wave theory. We infer that the small ordered moment is caused by the effect of exchange bond randomness arising from the site disorder of Sr and La ions.

\end{abstract}



\maketitle

\section{Introduction}
The quest for an experimental realization of a quantum disordered ground state (QDGS) is a frontier of condensed matter physics. In low-dimensional spin-$1/2$ magnets with competing exchange interactions, strong quantum fluctuations suppress magnetic ordering and lead to QDGSs such as the spin liquid state~\cite{Savary2017,Balents2010} and valence-bond-solid state~\cite{Read1989}. The spin-$1/2$ square-lattice Heisenberg antiferromagnet (SLHAF) with the nearest-neighbor (NN) $J_1$ and next-nearest-neighbor (NNN) $J_2$ exchange interactions, referred to as the $S\,{=}\,1/2\,J_1{-}J_2$ SLHAF, is a prototypical frustrated quantum magnet, whose ground state depends on the value of ${\alpha}\,{=}\,J_1/J_2$. The QDGS has been predicted to emerge in a critical range of $\alpha_1\,{<}\,J_1/J_2\,{<}\,{\alpha}_2$ with ${\alpha}_1\,{\simeq}\,0.4$ and ${\alpha}_2\,{\simeq}\,0.6$ ~\cite{Capriotti2001,Capriotti2000,Chandra1988,Dagotto1989,Darradi2008,Einarsson1995,Figueirido1990,Gong2014,Igarashi1993,Jiang2012,Mambrini2006,Mezzacapo2012,Read1991,Richter2010,Shannon2004,Shannon2006,Singh1999,Sirker2006,Sushkov2001,Zhang2003,Zhitomirsky1996}. However, no theoretical consensus has been achieved on the nature of the ground state, even for the existence of an excitation gap.
 On the other hand, the ground states for $J_1/J_2\,{<}\,{\alpha}_1$ and ${\alpha}_2\,{<}\,J_1/J_2$ are N\'{e}el antiferromagnetic (NAF) and columnar antiferromagnetic (CAF), respectively. Although many model materials such as the double perovskite Sr$_2$CuMO$_6$ (M\,{=}\,Mo, Te, and W)~\cite{Vasala2014a,Xu2017,Xu2018,Iwanaga1999,Babkevich2016,Vasala2014,Koga2016,Walker2016,Koga2014,Vasala2012} and vanadium compounds~\cite{Bombardi2004,Tsirlin2009,Carretta2009,Tsirlin2008,Tsirlin2010,Melzi2000,Melzi2001,Nath2008,Bossoni2011,Nath2009} have been reported, none of them are in the critical range. 
 
The effect of randomness in the exchange interaction in frustrated quantum magnets has been attracting much attention from the viewpoint of realistic routes to QDGSs~\cite{Kawamura2019}. Recently, it has been theoretically demonstrated that exchange randomness gives rise to a QDGS in the $S\,{=}\,1/2$ random $J_1{-}J_2$ SLHAF over a wide range of $J_1/J_2$~\cite{Liu2020,Liu2018,Uematsu2018}, while magnetic ordering survives in the case without frustration \cite{Laflorencie2006}. This QDGS is interpreted as a random singlet state~\cite{Lin2003,Fisher1994,Bhatt1982,Dasgupta1980} or valence-bond-glass state~\cite{Singh2010,Tarzia2008}. In real materials, there exist almost  unavoidable lattice disorders such as defects, dislocations, and site disorder of ions, which give rise to exchange randomness. Hence, it is necessary to investigate the effects of exchange randomness on the ground states in frustrated quantum magnets. Recently, the low-temperature magnetic properties of $B$-site ordered double perovskites Sr$_2$CuTe$_{1{-}x}$W$_x$O$_6$ have been actively investigated from the viewpoint of the $S\,{=}\,1/2$ random $J_1{-}J_2$ SLHAF~\cite{Hong2021,Mustonen2018_1,Mustonen2018_2,Watanabe2018,Yoon2021}. In the end member Sr$_2$CuTeO$_6$, $J_1$ is dominant, while $J_2$ is dominant in the other end member Sr$_2$CuWO$_6$. Thus, the random substitution of W$^{6+}$ ions for Te$^{6+}$ ions induces exchange randomness for $J_1$ and $J_2$ in the mixed system of Sr$_2$CuTe$_{1-x}$W$_x$O$_6$. It was found from muon spin rotation and relaxation ($\mu$SR)~\cite{Mustonen2018_2} and specific heat measurements~\cite{Watanabe2018} that magnetic ordering is strongly suppressed and the QDGS emerges in the wide range of $0.1\,{<}\,x\,{<}\,0.6$. However, the nature of the QDGS in Sr$_2$CuTe$_{1{-}x}$W$_x$O$_6$ is still under debate.

\begin{figure}[]
\begin{center}
\includegraphics[width=1\linewidth]{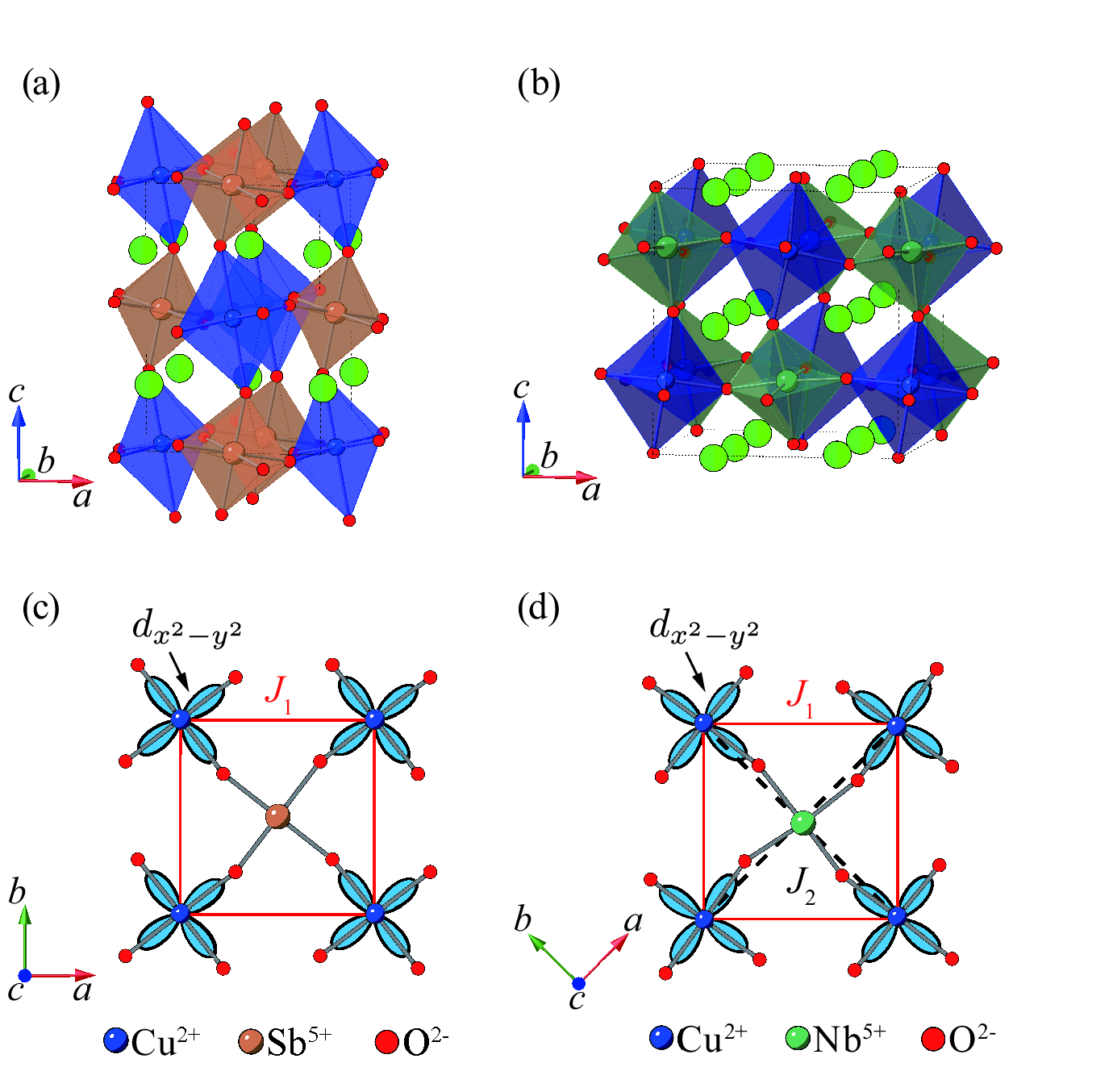}
\end{center}
\vspace{-15pt}\caption{Schematic view of the crystal structures of (a) SrLaCuSbO$_6$ (SLCSO) and (b) SrLaCuNbO$_6$ (SLCNO) with the $B$-site ordered double perovskite structure. Blue, brown, and green indicate CuO$_6$, SbO$_6$, and NbO$_6$ octahedra, respectively. Green and red spheres are Sr/La and oxide ions, respectively. (c) and (d) show the linkages of CuO$_6$ and Sb(Nb)O$_6$ octahedra in the $ab$-plane of SLCSO and SLCNO, respectively. The hole orbitals $d_{x^2-y^2}$ of Cu$^{2+}$ ions are shown in blue. The uniform NN interactions $J_1$ and NNN interactions $J_2$ are illustrated by red solid and black dashed lines in (d), respectively.}
\label{structure}
\end{figure}

In this paper, we report the magnetic properties of the $B$-site ordered double perovskites SrLaCuSbO$_6$ (SLCSO) and SrLaCuNbO$_6$ (SLCNO). Although the crystal structures of SLCSO and SLCNO were reported in previous studies~\cite{Attfield1992,West2011}, no studies on their magnetism have been published yet. At room temperature, SLCSO and SLCNO crystallize in a monoclinic structure $P2_1/n$ and a triclinic structure $P\bar{1}$, respectively~ \cite{Attfield1992,West2011}. In these compounds, CuO$_6$ and Sb(Nb)O$_{6}$ octahedra are arranged alternately in the $ab$-plane and Cu$^{2+}$ ions form a slightly distorted square lattice in the $ab$-plane as shown in Figs.~\ref{structure}\,(a) and (b). CuO$_6$ octahedra are elongated approximately parallel to the $c$-axis owing to the Jahn--Teller effect. Consequently, the hole orbitals $d_{x^2-y^2}$ of Cu$^{2+}$ ions with spin-1/2 are spread in the $ab$-plane, as shown in Figs.~\ref{structure}\,(c) and (d), which leads to good two-dimensionality. While super-exchange interactions are expected between the NN and NNN spins via nonmagnetic Sb(Nb)O$_6$ octahedra, the disorder of Sr$^{2+}$ and La$^{+3}$ ions produces weak randomness in the magnitude of exchange interaction. Therefore, it is considered that SLCSO and SLCNO are described as the $S\,{=}\,1/2$ $J_1{-}J_2$ SLHAF with weak bond randomness.

\section{Experimental details}
Polycrystalline samples of SLCSO and SLCNO were synthesized via a conventional solid-state method using SrCO$_{3}$~(99.9\,\%), La$_{2}$O$_{3}$~(99.9\,\%), CuO~(99.99\,\%), Sb$_{2}$O$_{5}$~(99.99\,\%), and Nb$_{2}$O$_{5}$~(99.99\,\%) as starting materials. A stoichiometric mixture of the starting materials was ground well with an agate mortar and fired at 1000\,$^{\circ}$C for 24\,h. The powder samples obtained were then reground, pelletized, and calcined twice at 1200\,$^{\circ}$C for 24\,h. The phase purities of the samples were confirmed by powder X-ray diffraction measurements using a MiniFlex\,II diffractometer (Rigaku) with Cu\,$K_{\alpha}$ radiation at room temperature. 

Neutron powder diffraction (NPD) measurements were performed at 2.6, 20, and 300\,K using the SuperHRPD(BL08) time-of-flight diffractometer installed at the Material and Life Science Facility (MLF) at J-PARC, Japan~\cite{Torii2011,Torii2014}. The crystal structures of SLCSO and SLCNO were refined using the high-resolution NPD profiles obtained from the backscattering (BS) bank by Rietveld analysis with the Fullprof program~\cite{Rodriguez1993}. For the magnetic structure analysis, the diffraction data were collected from the 90 degree (QA) bank and low-angle (LA) bank. The magnetic structures of SLCSO and SLCNO were determined by irreducible representation analysis using the SARAh program~\cite{WILLS2000}.

X-band (9.44 GHz) electron spin resonance (ESR) spectra of SLCSO and SLCNO were collected with a Bruker spectrometer at room temperature to estimate the $g$-factor. ESR data were analyzed using the EasySpin software package \cite{STOLL2006}. 

Magnetization $M(H,T)$ measurements on SLCSO and SLCNO were performed in the temperature range of $1.8\,{\leq}\,T\,{\leq}\,300\,{\rm K}$ at applied magnetic fields of $0.1\,{\leq}\,\mu_{0}H\,{\leq}\,7\,{\rm T}$ using a superconducting quantum interference device (SQUID) magnetometer (MPMS XL, Quantum Design). Specific heat $C(H, T)$ measurements  on SLCSO and SLCNO were carried out down to 1.8 K at magnetic fields of up to 9 T using a physical property measurement system (PPMS, Quantum Design) by the relaxation method. 
Nuclear magnetic resonance (NMR) measurements were performed on SLCSO and SLCNO using a 16 T superconducting magnet in the temperature range between 1.8 and 20 K.

$^{121}$Sb and $^{93}$Nb NMR measurements were performed on powder samples of SLCSO and SLCNO, respectively, using a 12 T superconducting magnet in the temperature range between 4 and 30 K.  Spectra were obtained by recording the spin-echo amplitude against a slowly varying applied field~\cite{Oosawa2009}.  The measurement field region was restricted to the central transition part of the electric quadrupolar powder pattern for $I\,{=}\,7/2$ ($^{121}$Sb) and $I\,{=}\,5/2$ ($^{93}$Nb) nuclei. 
The longitudinal nuclear spin relaxation rate was obtained by the saturation recovery method using a pulse train~\cite{Matsui2017}. The recovery of nuclear spin magnetization was traced after saturation until the difference from the thermal equilibrium value became less than 1\%. The relaxation time $T_1$ was evaluated by fitting the observed recovery curves to the conventional theoretical formulas of $0.363e^{-45 \tau/T_1}+0.192e^{-28\tau/T_1}+0.153e^{-15 \tau/T_1}+0.140e^{-6\tau/T_1}+0.152e^{-\tau/T_1}$ for $^{93}$Nb and $0.029e^{-\tau/T_1}+0.178e^{-6\tau/T_1}+0.793e^{-6\tau/T_1}$ for $^{121}$Sb for the central transition of nuclear spins with electric quadrupole interaction~\cite{Narath1967, Lue2011}.
\begin{table*}[]
\caption{Structural parameters of SrLaCuSbO$_6$ obtained by Rietveld analysis of neutron power diffraction spectra at 300 and 20\,K.}
\label{slcs_str}
\begin{ruledtabular}
\begin{tabular}{llllllllllllll}
&\multicolumn{6}{c}{300 K}&&&\multicolumn{5}{c}{20 K}\\\hline
&\multicolumn{6}{l}{\hspace{3mm}Space group: $P2_1/n$ (No.14)}&&&\multicolumn{5}{l}{\hspace{3mm}Space group: $P2_1/n$ (No.14)}\rule[0mm]{0mm}{4mm}\vspace{0.6mm}\\
&\multicolumn{6}{l}{\hspace{3mm}$a\,{=}\, 5.51406(2)$\,{\AA},\ $b\,{=}\,5.510057(20)$\,{\AA},\ $c\,{=}\,8.39673(3)$\,{\AA}}&&&\multicolumn{5}{l}{\hspace{3mm}$a\,{=}\,5.504498(19)$\,{\AA},\ $b\,{=}\,5.50362(17)$\,{\AA},\ $c\,{=}\,8.38456(2)$\,{\AA}}\vspace{0.6mm}\\
&\multicolumn{6}{l}{\hspace{3mm}$\alpha\,{=}\,90^\circ,\ \beta\,{=}\,90.4888(3)^\circ,\ \gamma\,{=}\,90^\circ$}&&&\multicolumn{5}{l}{\hspace{3mm}$\alpha\,{=}\,90^\circ,\ \beta\,{=}\,90.5138(3)^\circ,\ \gamma\,{=}\,90^\circ$}\vspace{0.6mm}\\
&\multicolumn{6}{l}{\hspace{3mm}$R_{\rm p}\,{=}\,9.33\,\%,\ R_{\rm wp}\,{=}\,11.7\,\%,\ R_{\rm e}\,{=}\,2.26\,\%$}&&&\multicolumn{5}{l}{\hspace{3mm}$R_{\rm p}\,{=}\,8.80\,\%,\ R_{\rm wp}\,{=}\,9.04\,\%,\ R_{\rm e}\,{=}\,0.88\,\%$}\vspace{0.6mm}\\
&\multicolumn{6}{l}{\hspace{3mm}$R_{\rm B}\,{=}\,3.89\,\%,\ R_{\rm F}\,{=}\,3.08\,\%$}&&&\multicolumn{5}{l}{\hspace{3mm}$R_{\rm B}\,{=}\,4.60\,\%,\ R_{\rm F}\,{=}\,3.22\,\%$}\vspace{0.6mm}\\\hline
Atom & Site &\ $x$ &\ $y$ &\ $z$ & $g$ & $B_{\rm iso}$ &&&\ $x$ &\ $y$ &\ $z$ & $g$ & $B_{\rm iso}$\rule[0mm]{0mm}{4mm}\\   
Sr/La&	4$e$	&\	0.0109(2)	&\	0.0207(2)	&	0.24859(14)	&	0.5/0.5	&	0.61(2)	&&&\	0.0109(19)	&\	0.23366(20)	&	0.24861(12)&	0.5/0.5	&	0.29(2)	\\
Cu	&	2$c$	&\	1/2			&\	0			&	1/2			&	1		&	0.34(3)	&&&\	1/2			&\	0			&	1/2			&	1		&	0.17(2)	\\
Sb	&	2$d$	&\	1/2			&\	0			&	0			&	1		&	0.11(3)	&&&\	1/2			&\	0			&	0			&	1		&	0.06(3)	\\
O1	&	4$e$	&\	0.2815(3)	&\	0.2808(3)	&	0.0391(2)	&	1		&	0.84(3)	&&&\	0.2829(3)	&\	0.2834(3)	&	0.04046(18)	&	1		&	0.70(2)	\\
O2	&	4$e$	&\	0.2218(3)	&	$-$0.2190(3)	&	0.038107(19)	&	1		&	0.84(3)	&&&\	0.2203(3)	&	$-$0.2184(3)	&	0.03937(17)	&	1		&	0.67(3)	\\
O3	&	4$e$	&	$-$0.0695(2)	&\	0.4919(3)	&	0.27144(16)	&	1		&	0.65(2)	&&&	$-$0.0724(2)	&\	0.4909(3)	&	0.27164(15)	&	1		&	0.56(3)	\\
\end{tabular}
\end{ruledtabular}
\end{table*}
\begin{figure}[]
\begin{center}
\includegraphics[width=1\linewidth]{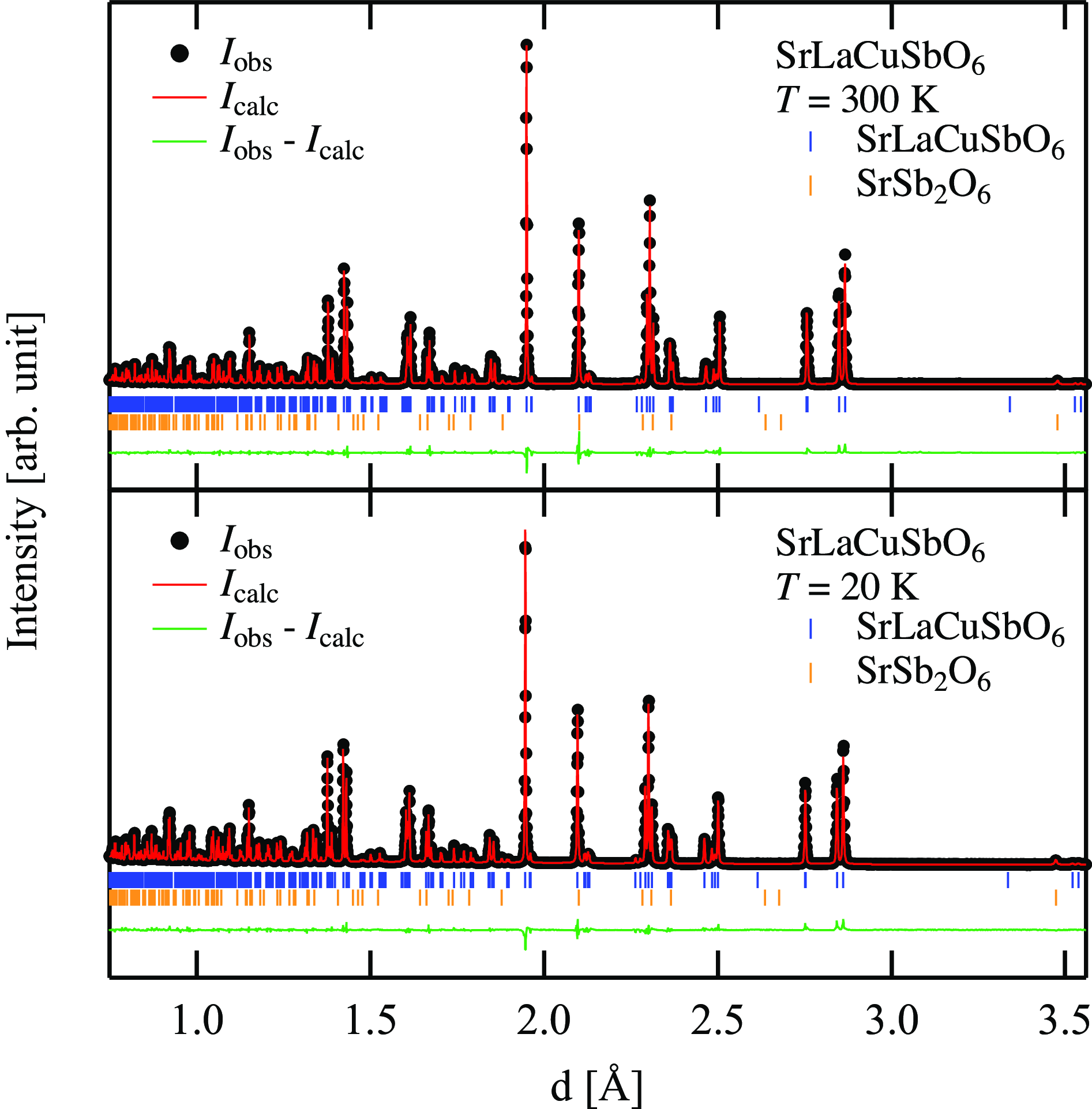}
\end{center}
\vspace{-15pt}\caption{NPD patterns of SLCSO collected at 300 and 20\,K, and the results of Rietveld fitting (red lines). Green curves show the differences between the observed and calculated intensities. A minor 0.96(3)\,wt\% nonmagnetic impurity of SrSb$_2$O$_6$ was found.}
\label{npd_rt20K_SLCS}
\end{figure}
\begin{figure}[]
\begin{center}
\includegraphics[width=1\linewidth]{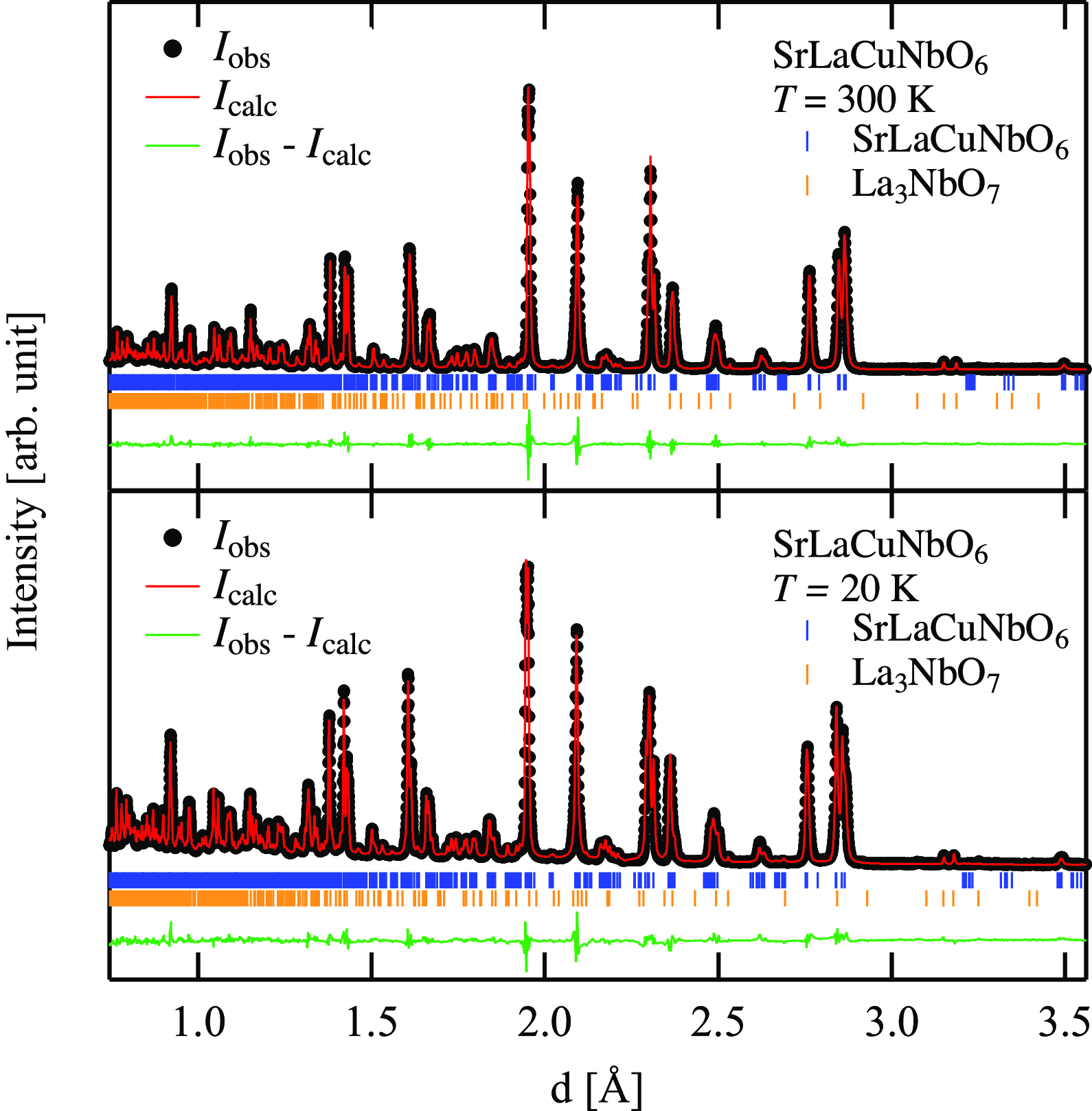}
\end{center}
\vspace{-15pt}\caption{NPD patterns of SLCNO collected at 300 and 20\,K, and the results of Rietveld fitting (red lines). Green curves show the differences between the observed and calculated intensities. A minor 1.0(1) wt\% nonmagnetic impurity of La$_3$NbO$_7$ was found.}
\label{npd_rt20K_SLCN}
\end{figure}

\section{Results}
\subsection{Crystal structure}
Figure~\ref{npd_rt20K_SLCS} shows NPD patterns for SLCSO collected from the BS bank at 300 and 20\,K, and the results of Rietveld analysis. The analysis was based on the $P2_1/n$ structure reported by Attfield {\it et al.}~\cite{Attfield1992}. It was found that the SLCSO sample contains a minor 0.96(3) wt\% nonmagnetic impurity of SrSb$_2$O$_6$. Also, there is no structural transition and SLCSO retains the $P2_1/n$ structure down to 3.8 K. The structural parameters obtained by Rietveld analysis are listed in Table~\ref{slcs_str}. 
\begin{table*}[]
\caption{Structural parameters of SrLaCuNbO$_6$ obtained by Rietveld analysis of neutron power diffraction spectra at 300 and 20\,K.}
\label{slcn_str}
\begin{ruledtabular}
\begin{tabular}{lllllllllllllll}
&\multicolumn{6}{c}{300 K}&&&&\multicolumn{5}{c}{20 K}\\\hline
&\multicolumn{6}{l}{\hspace{3mm}Space group: $P{\bar 1}$ (No.2)}&&&&\multicolumn{5}{l}{\hspace{3mm}Space group: $P{\bar 1}$ (No.2)}\rule[0mm]{0mm}{4mm}\vspace{0.6mm}\\
&\multicolumn{6}{l}{\hspace{3mm}$a\,{=}\,7.80121(6)$\,{\AA},\ $b\,{=}\,7.82001(7)$\,{\AA},\ $c\,{=}\,8.36915(8)$\,{\AA}}&&&&\multicolumn{5}{l}{\hspace{3mm}$a\,{=}\,7.78131(5)$\,{\AA},\ $b\,{=}\,7.80553(5)$\,{\AA},\ $c\,{=}\,8.36347(7)$\,{\AA}}\vspace{0.6mm}\\
&\multicolumn{6}{l}{\hspace{3mm}$\alpha\,{=}\,89.6533(11)^\circ,\ \beta\,{=}\,89.6630(9)^\circ,\ \gamma\,{=}\,89.9235(9)^\circ$}&&&&\multicolumn{5}{l}{\hspace{3mm}$\alpha\,{=}\,89.5094(8)^\circ,\ \beta\,{=}\,89.6578(7)^\circ,\ \gamma\,{=}\,89.9049(7)^\circ$}\vspace{0.6mm}\\
&\multicolumn{6}{l}{\hspace{3mm}$R_{\rm p}\,{=}\,9.02\,\%,\ R_{\rm wp}\,{=}\,10.1\,\%,\ R_{\rm e}\,{=}\,2.44\,\%$}&&&&\multicolumn{5}{l}{\hspace{3mm}$R_{\rm p}\,{=}\,10.3\,\%,\ R_{\rm wp}\,{=}\,8.13\,\%,\ R_{\rm e}\,{=}\,0.745\,\%$}\vspace{0.6mm}\\
&\multicolumn{6}{l}{\hspace{3mm}$R_{\rm B}\,{=}\,3.38\,\%,\ R_{\rm F}\,{=}\,5.09\,\%$}&&&&\multicolumn{5}{l}{\hspace{3mm}$R_{\rm B}\,{=}\,5.84\,\%,\ R_{\rm F}\,{=}\,5.88\,\%$}\vspace{0.6mm}\\\hline
Atom & Site &\ $x$ &\ $y$ &\ $z$ & $g$ & $B_{\rm iso}$ &&&&\ $x$ &\ $y$ &\ $z$ & $g$ & $B_{\rm iso}$\rule[0mm]{0mm}{4mm}\\   
Sr1/La1&	2$i$	&\	0.2499(16)	&\	0.4855(15)	&\	0.4960(15)	&	0.5/0.5	&	1.09(3)	&&&&\	0.2513(19)	&\	0.483(2)	&\	0.494(2)	&	0.5/0.5	&	0.30(3)	\\
Sr2/La2	&	2$i$	&\	0.2525(16)	&	$-$0.0150(15)	&\	0.4916(14)	&	0.5/0.5	&	1.09(3)	&&&&\	0.2522(17)	&	$-$0.019(2)	&\	0.483(3)	&	0.5/0.5	&	0.30(3)	\\
Sr3/La3	&	2$i$	&\	0.2514(15)	&\	0.4822(15)	&	$-$0.0195(12))	&	0.5/0.5	&	1.09(3)	&&&&\	0.2480(18)	&\	0.484(2)	&	$-$0.018(3)	&	0.5/0.5	&	0.30(3)	\\
Sr4/La4	&	2$i$	&\	0.7529(15)	&\	0.0155(15)	&\	0.0136(14)	&	0.5/0.5	&	1.09(3)	&&&&\	0.7495(18)	&\	0.017(2)	&\	0.009(2)	&	0.5/0.5	&	0.30(3)	\\
Nb1	&	2$i$	&\	0.004(2)	&\	0.252(2)	&\	0.7549(17)	&	1	&	0.24(6)	&&&&\	0.005(2)	&\	0.253(2)	&\	0.755(2)	&	1	&	0.03(3)	\\
Nb2	&	2$i$	&\	0.505(2)	&\	0.255(2)	&\	0.2515(18)	&	1	&	0.24(6)	&&&&\	0.499(2)	&\	0.2618(16)	&\	0.2581(16)	&	1	&	0.03(3)	\\
Cu1	&	2$i$	&\	0.003(2)	&\	0.253(2)	&\	0.2534(19)	&	1	&	0.61(5)	&&&&\	0.005(2)	&\	0.253(2)	&\	0.253(3)	&	1	&	0.03(3)	\\
Cu2	&	2$i$	&\	0.5002(20)	&\	0.256(2)	&\	0.7563(17)	&	1	&	0.61(5)	&&&&\	0.504(3)	&\	0.255(2)	&\	0.750(3)	&	1	&	0.03(3)	\\
O11	&	2$i$	&\	0.0400(18)	&\	0.002(2)	&\	0.2255(17)	&	1	&	1.11(2)	&&&&\	0.042(2)	&\	0.0103(19)	&\	0.220(3)	&	1	&	0.57(2)	\\
O12	&	2$i$	&	$-$0.0501(18)	&\	0.2737(18)	&	$-$0.0145(15)	&	1	&	1.11(2)	&&&&	$-$0.047(2)	&\	0.277(2)	&	$-$0.023(2)	&	1	&	0.57(2)	\\
O13	&	2$i$	&	$-$0.0311(18)	&\	0.500(2)	&\	0.2893(18)	&	1	&	1.11(2)	&&&&	$-$0.033(2)	&\	0.502(2)	&\	0.287(3)	&	1	&	0.57(2)	\\
O14	&	2$i$	&\	0.0356(18)	&\	0.2113(18)	&\	0.5187(15)	&	1	&	1.11(2)	&&&&\	0.041(2)	&\	0.209(2)	&\	0.514(2)	&	1	&	0.57(2)	\\
O21	&	2$i$	&\	0.2482(20)	&\	0.2936(18)	&\	0.2265(15)	&	1	&	1.11(2)	&&&&\	0.248(3)	&\	0.292(3)	&\	0.221(3)	&	1	&	0.57(2)	\\
O22	&	2$i$	&\	0.250(2)	&\	0.2248(19)	&\	0.7928(17)	&	1	&	1.11(2)	&&&&\	0.252(3)	&\	0.225(3)	&\	0.798(3)	&	1	&	0.57(2)	\\
O23	&	2$i$	&\	0.246(2)	&\	0.7257(20)	&\	0.2987(17)	&	1	&	1.11(2)	&&&&\	0.254(3)	&\	0.723(3)	&\	0.297(4)	&	1	&	0.57(2)	\\
O24	&	2$i$	&\	0.250(2)	&\	0.7857(18)	&\	0.7166(16)	&	1	&	1.11(2)	&&&&\	0.250(3)	&\	0.792(3)	&\	0.723(3)	&	1	&	0.57(2)	\\
O31	&	2$i$	&\	0.4625(19)	&\	0.003(2)	&\	0.2187(19)	&	1	&	1.11(2)	&&&&\	0.466(3)	&	$-$0.002(3)	&\	0.211(3)	&	1	&	0.57(2)	\\
O32	&	2$i$	&\	0.5292(19)	&\	0.283(2)	&\	0.0195(17)	&	1	&	1.11(2)	&&&&\	0.5230(17)	&\	0.2810(18)	&\	0.022(2)	&	1	&	0.57(2)	\\
O33	&	2$i$	&\	0.4530(18)	&\	0.209(2)	&\	0.4801(17)	&	1	&	1.11(2)	&&&&\	0.4491(16)	&\	0.2044(18)	&\	0.4739(17)	&	1	&	0.57(2)	\\
O34	&	2$i$	&\	0.5301(18)	&\	0.499(2)	&\	0.2934(19)	&	1	&	1.11(2)	&&&&\	0.534(3)	&\	0.497(3)	&\	0.285(3)	&	1	&	0.57(2)	\\
\end{tabular}
\end{ruledtabular}
\end{table*}

Figure~\ref{npd_rt20K_SLCN} shows NPD patterns for SLCNO collected from the BS bank at 300 and 20 K, and the results of Rietveld analysis. The analysis was performed on the basis of a structural model with the space group $P\bar{1}$ reported by West and Davies~\cite{West2011}. Structural parameters of SrLaCuTaO$_6$~\cite{West2011}, which is isostructural with SLCNO, were chosen as initial parameters of Rietveld refinement, and the isotropic atomic displacement parameter $B_{\rm iso}$ was fixed for each type of atom. A minor 1.0(1)\,wt\% nonmagnetic impurity of La$_3$NbO$_7$ was found. The structural parameters obtained by Rietveld analysis are listed in Table~\ref{slcn_str}. 

It was confirmed from the present structural refinement that all the CuO$_6$ octahedra are elongated approximately along the $c$-axis owing to the Jahn--Teller effect in both compounds. The hole orbital $d_{x^2-y^2}$, which gives the lowest orbital level of the Cu$^{2+}$ ion, is spread in the $ab$-plane, as shown in Figs.~\ref{structure}\,(c) and (d). Hence, the super-exchange interactions mediated via O$^{2-}$ ions in the $ab$-plane must be much larger than those along the $c$-axis, which leads to the good two-dimensionality in SLCSO and SLCNO. The crystal lattices of SLCSO and SLCNO are nearly tetragonal, as shown in Tables~\ref{slcs_str} and \ref{slcn_str}; thus, the NN and NNN exchange interactions are expected to be nearly uniform in both compounds. 

\subsection{Electron spin resonance}
\begin{figure}[t]
\begin{center}
\includegraphics[width=0.9\linewidth]{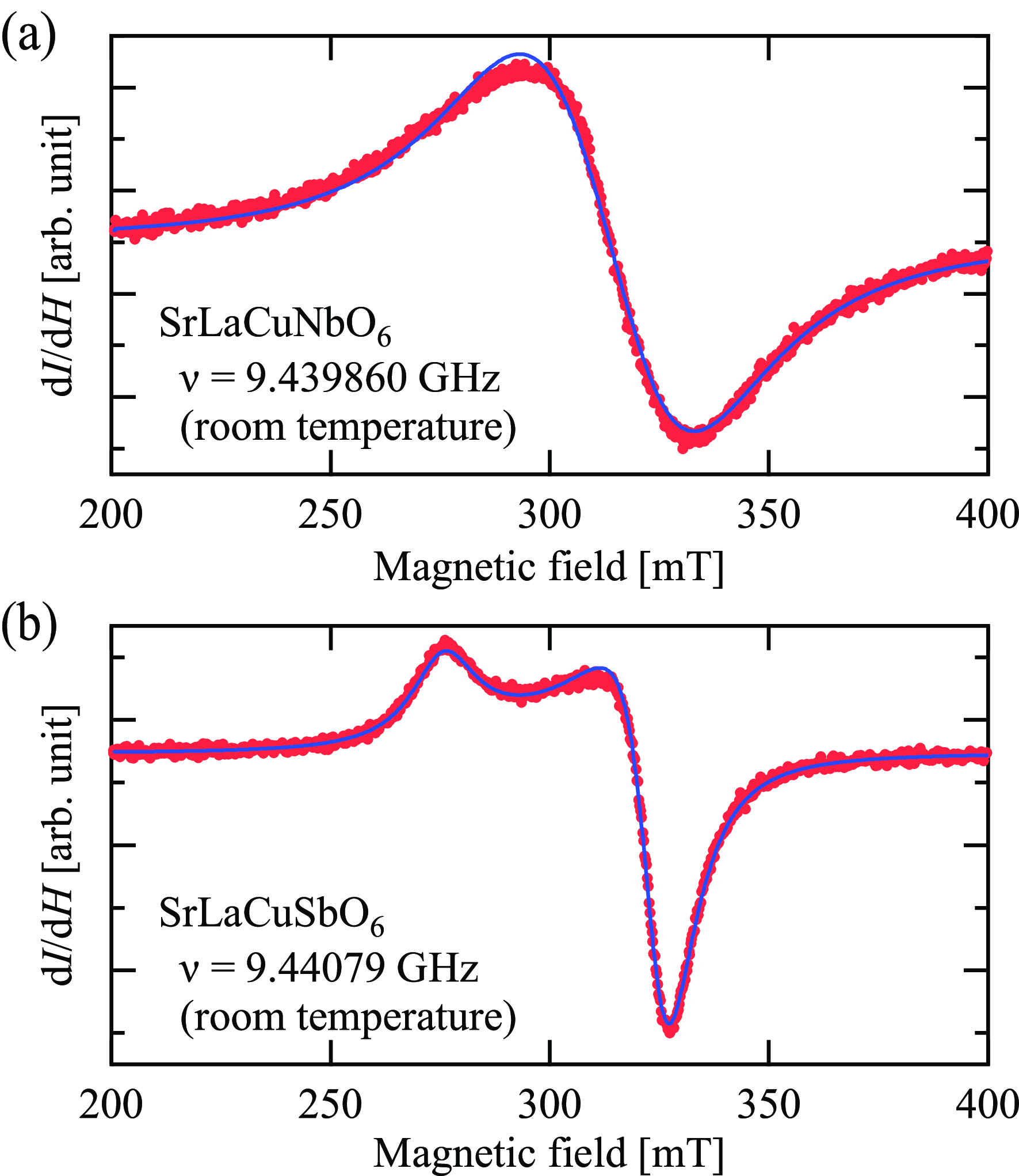}
\end{center}
\vspace{-15pt}\caption{Field derivative of the ESR absorption intensity ${\rm d}I/{\rm d}H$ for (a) SLCSO and (b) SLCNO. Blue solid lines are fits using two pseudo-Voigt functions with the anisotropic $g$-factors given in the text.}
\label{esr}
\end{figure}

We performed ESR measurements on SLCSO and SLCNO to estimate the $g$-factors. Figures~\ref{esr}\,(a) and (b) show the field derivative of the ESR absorption intensity $dI/dH$ for SLCSO and SLCNO, respectively, measured at room temperature using an X-band spectrometer. The shape of the ESR spectrum of SLCSO is typical of the powder spectrum of Cu$^{2+}$ ions in an elongated octahedral environment with two principal values of the $g$-factors $g_\parallel$ and $g_{\perp}$ for magnetic fields parallel and perpendicular to the elongated axis, respectively. Fitting a superposition of field derivatives of two pseudo-Voigt functions to the ESR spectrum for SLCSO, we obtain $g_{\parallel}\,{=}\,2.437$ and $g_{\perp}\,{=}\,2.076$ for SLCSO. The blue solid line in Fig.~\ref{esr}\,(a) indicates the fit. These $g$-factors are typical for copper (II) oxides with elongated CuO$_6$ octahedra. 

The ESR spectrum of SLCNO is symmetric, but its linewidth is much broader than that of SLCSO. We infer that line broadening arises from the additional magnetic anisotropy such as the antisymmetric exchange interaction of the Dzyaloshinskii--Moriya type, which is induced by the low symmetry of the crystal lattice. We obtain $g_{\parallel}\,{=}\,2.259$ and $g_{\perp}\,{=}\,2.100$ from the same analysis as that applied to the ESR spectrum of SLCSO. The average of the $g$-factors for SLCNO, $g_{\rm avg}\,{=}\,2.15$, is smaller than $g_{\rm avg}\,{=}\,2.20$ for SLCSO, which is consistent with the Curie constants of both compounds shown below. Note that the accuracy of the $g$-factors for SLCNO will be significantly worse than that for SLCSO, because the ESR spectrum of SLCNO is not well reproduced by the fit with a single component of the derivative pseudo-Voigt function and the linewidth of SLCNO is much broader than that of SLCSO. However, the average of the $g$-factors for SLCNO is considered to be close to the true value. The CuO$_6$ octahedron in SLCNO is tetragonally elongated and the Cu$^{2+}$ ion is located almost at the center of the octahedron. For SLCNO, there are two kinds of CuO$_6$ octahedron, which are elongated approximately along the $c$-axis. In both octahedra, Cu$^{2+}$ ions are off-centered. We infer that the difference in the local environment of Cu$^{2+}$ ion gives rise to the difference in the values of $g_{\rm avg}$ between these two compounds.

\subsection{Magnetic susceptibility and magnetization}
\begin{figure}[h!]
\begin{center}
\includegraphics[width=1\linewidth]{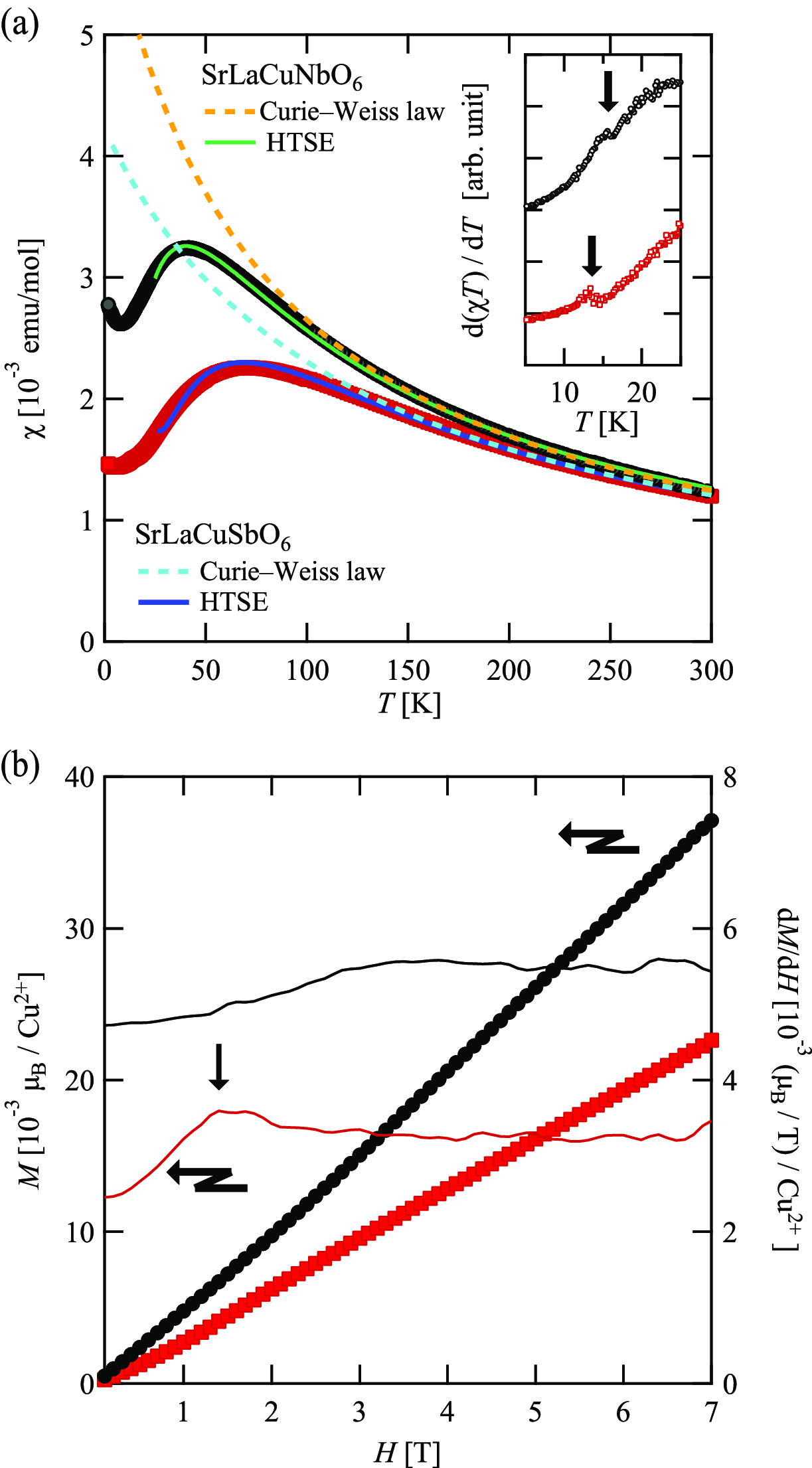}
\end{center}
\vspace{-15pt}\caption{(Color online) (a) Temperature dependence of magnetic susceptibilities $\chi$ of SLCSO (red squares) and SLCNO (black circles) measured at ${\mu}_0H\,{=}\,0.1$\,T. The green and blue solid curves are fits by the HTSEs of the $S\,{=}\,1/2$ $J_1{-}J_2$ SLHAF for SLCNO and SLCSO, respectively. The orange and light blue dashed curves are fits by the Curie--Weiss law for SLCNO and SLCSO in the temperature range of 150-300\,K, respectively. The inset shows the temperature dependences of $d({\chi}T)/dT$ for SLCSO and SLCNO between 5 and 25\,K. Downward arrows indicate small humps suggestive of magnetic phase transitions at $T_{\rm N}\,{=}\,13.6\,$K for SLCSO and at $T_{\rm N}\,{=}\,15.7\,$K for SLCNO. (b) Magnetization curves of SLCSO and SLCNO measured at 1.8\,K. The solid curves show the field derivatives ${\rm d}M/{\rm d}H$. The downward arrow indicates a small bend anomaly in the ${\rm d}M/{\rm d}H$ data of SLCNO, which is suggestive of a spin-flop transition.}
\label{mtmh}
\end{figure}

The temperature dependence of the magnetic susceptibilities of SLCSO and SLCNO measured at $H\,{=}\,0.1\,$T is shown in Fig.~\ref{mtmh}\,(a). With decreasing temperature, the magnetic susceptibility ($\chi$) exhibits a broad maximum at $T_{\rm max}\,{=}\,71\,$K for SLCSO and at $T_{\rm max}\,{=}\,44\,$K for SLCNO. The broad maximum is attributable to the evolution of magnetic short-range correlations, which is characteristic of two-dimensional SLHAFs~\cite{Jongh1974,Kim1998,Rosner2003}. Below $T_{\rm max}$, small hump anomalies indicative of magnetic ordering are observed in ${\rm d}({\chi}T)/{\rm d}T$ at $T_{\rm N}\,{=}\,13.6\,$K for SLCSO and at  $T_{\rm N}\,{=}\,15.7\,$K for SLCNO, as shown in the inset of Fig.~\ref{mtmh}\,(a). Above 150\,K, the magnetic susceptibilities of both compounds are described by the Curie--Weiss law $\chi\,{=}\,C/(T\,{-}\,\Theta)$ with $C\,{\simeq}\,0.51\,{\rm emu\,K\,mol^{-1}}$ and $\Theta\,{\simeq}\,{-}\,120$\,K for SLCSO, and  $C\,{\simeq}\,0.47\,{\rm emu\,K\,mol^{-1}}$ and $\Theta\,{\simeq}\,\,{-}\,77$\,K for SLCNO. The large Weiss constants $\Theta$ indicate that the dominant exchange interaction is antiferromagnetic and large. Details of exchange parameters will be discussed in Sec. \ref{discussion}.

Figure~\ref{mtmh}\,(b) shows the magnetic field dependence of magnetization $M$ and its field derivative ${\rm d}M/{\rm d}H$ for SLCSO and SLCNO at 1.8\,K. A small cusp anomaly of ${\rm d}M/{\rm d}H$ is observed around 1.4\,T for SLCSO, which indicates a spin-flop transition. Since ordered spins lie in the $ab$-plane as shown in Sec. \ref{mag_struc_SLCSO}, the spin-flop transition is caused by easy-axis anisotropy in the $ab$-plane. The spin-flop transition was not detected in SLCNO powder, which suggests that the in-plane anisotropy is negligible in SLCNO. 

\subsection{Specific heat measurement}
\begin{figure}[]
\begin{center}
\includegraphics[width=0.95\linewidth]{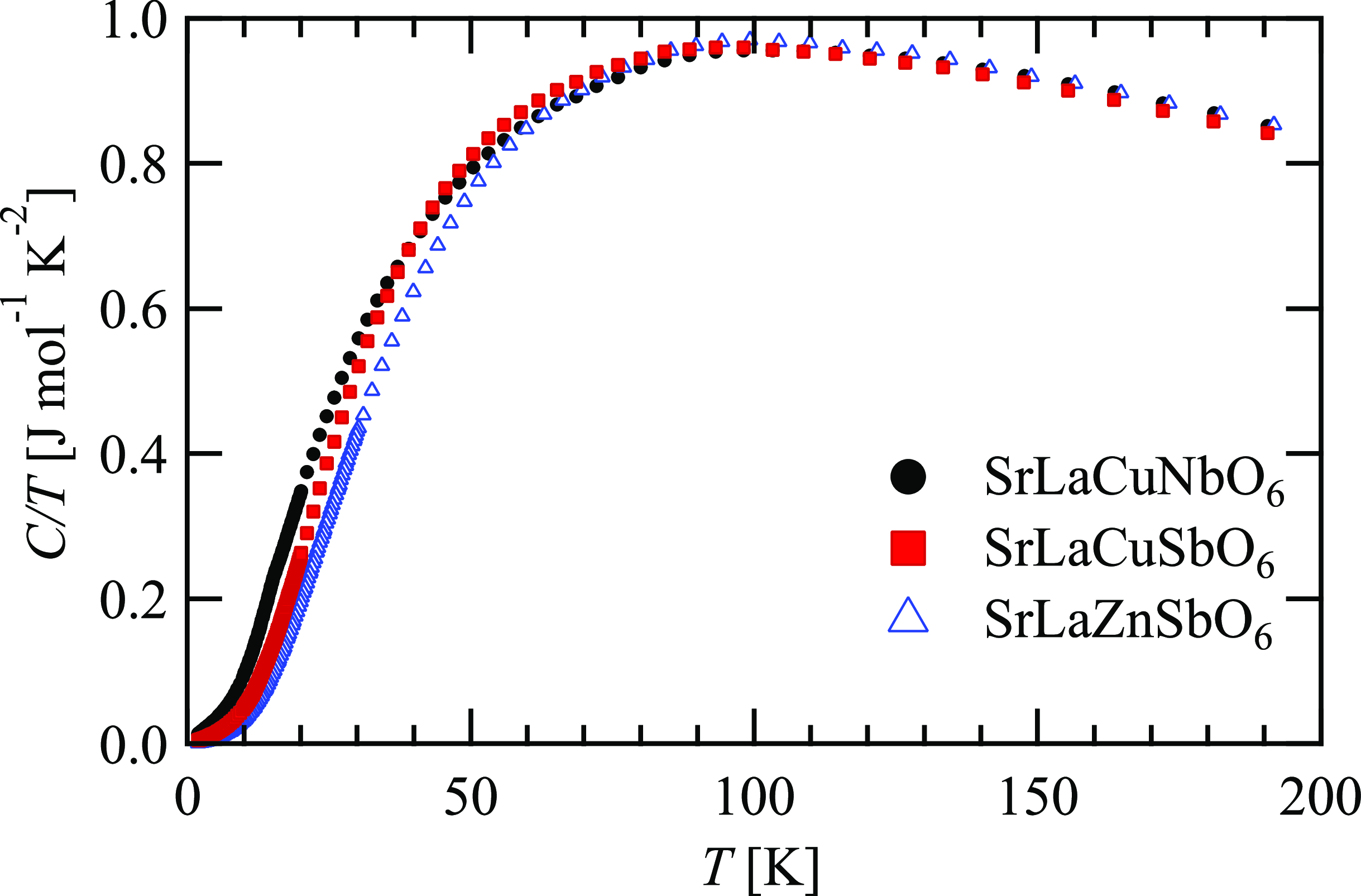}
\end{center}
\vspace{-15pt}\caption{(Color online) Temperature dependence of total specific heat divided by temperature $C/T$ in SLCSO, SLCNO, and nonmagnetic SrLaZnSbO$_6$ powders measured at zero magnetic field.}
\label{hc1}
\end{figure}

\begin{figure}[]
\begin{center}
\includegraphics[width=1\linewidth]{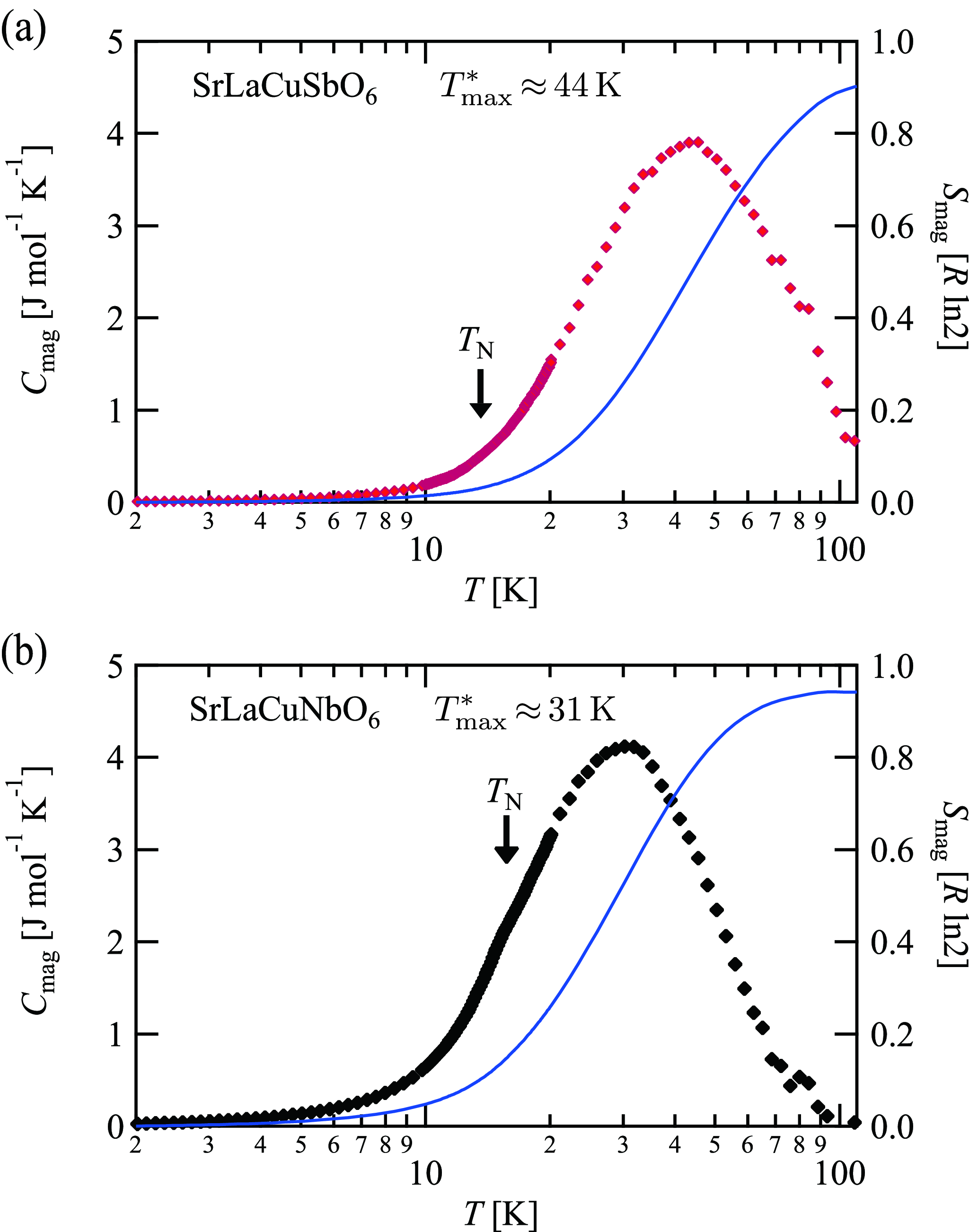}
\end{center}
\vspace{-15pt}\caption{(Color online) Temperature dependence of $C_{\rm mag}$ for (a) SLCSO and (b) SLCNO at zero magnetic field. Solid curves represent the magnetic entropy $S_{\rm mag}$. Vertical arrows indicate ordering temperatures $T_{\rm N}$ determined from the magnetic susceptibility data. }
\label{hc2}
\end{figure}

Figure~\ref{hc1} shows the temperature dependence of the total specific heat divided by temperature $C/T$ in SLCSO, SLCNO, and nonmagnetic SrLaZnSbO$_6$ measured at zero magnetic field. No anomaly indicative of a phase transition was observed down to 1.8 K in these three compounds. We extracted the magnetic contribution of the specific heat $C_{\rm mag}$ by subtracting the lattice contribution from the raw data. The lattice contribution was evaluated using the nonmagnetic counterpart SrLaZnSbO$_6$, which has a similar crystal structure to  SLCSO and SLCNO. Figures~\ref{hc2}\,(a) and (b) show the temperature dependence of $C_{\rm mag}$ at zero magnetic field. 

The magnetic entropy $S_{\rm mag}(T)$ was obtained by integrating $C_{\rm mag}/T$ over the interval 1.8\,K to $T$. The obtained $S_{\rm mag}(T)$ is shown on the right vertical axis of Figs.~\ref{hc2}\,(a) and (b). The magnetic entropies at 100 K are $S_{\rm mag}\,{\simeq}\,0.89R\,\ln 2$ and $0.94R\,\ln2$ for SLCSO and SLCNO, respectively, which are consistent with the theoretically expected value of $R\,\ln2$ for the $S\,{=}\,1/2$ systems. 

Small kinks were observed in $C_{\rm mag}$ at temperatures very close to the ordering temperatures $T_{\rm N}$ determined from the magnetic susceptibility data for both compounds, indicative of the magnetic ordering, as shown in Figs.~\ref{hc2}\,(a) and (b). These anomalies are small because the magnetic entropy is considerably decreased by the development of short-range correlations above $T_{\rm N}$ as $S(T_{\rm N})\,{\simeq}\,0.03R\,\ln2$ and $0.11R\,\ln2$ for SLCSO and SLCNO, respectively. A similar behavior of specific heat was also found in other $S\,{=}\,1/2$ $J_1\,{-}\,J_2$ SLHAFs Sr$_2$CuMO$_6$ (M\,{=}\,Te and W)~\cite{Mustonen2018_1}.


\subsection{Magnetic structure of SrLaCuSbO$_6$}\label{mag_struc_SLCSO}
\begin{figure}[]
\begin{center}
\includegraphics[width=1\linewidth]{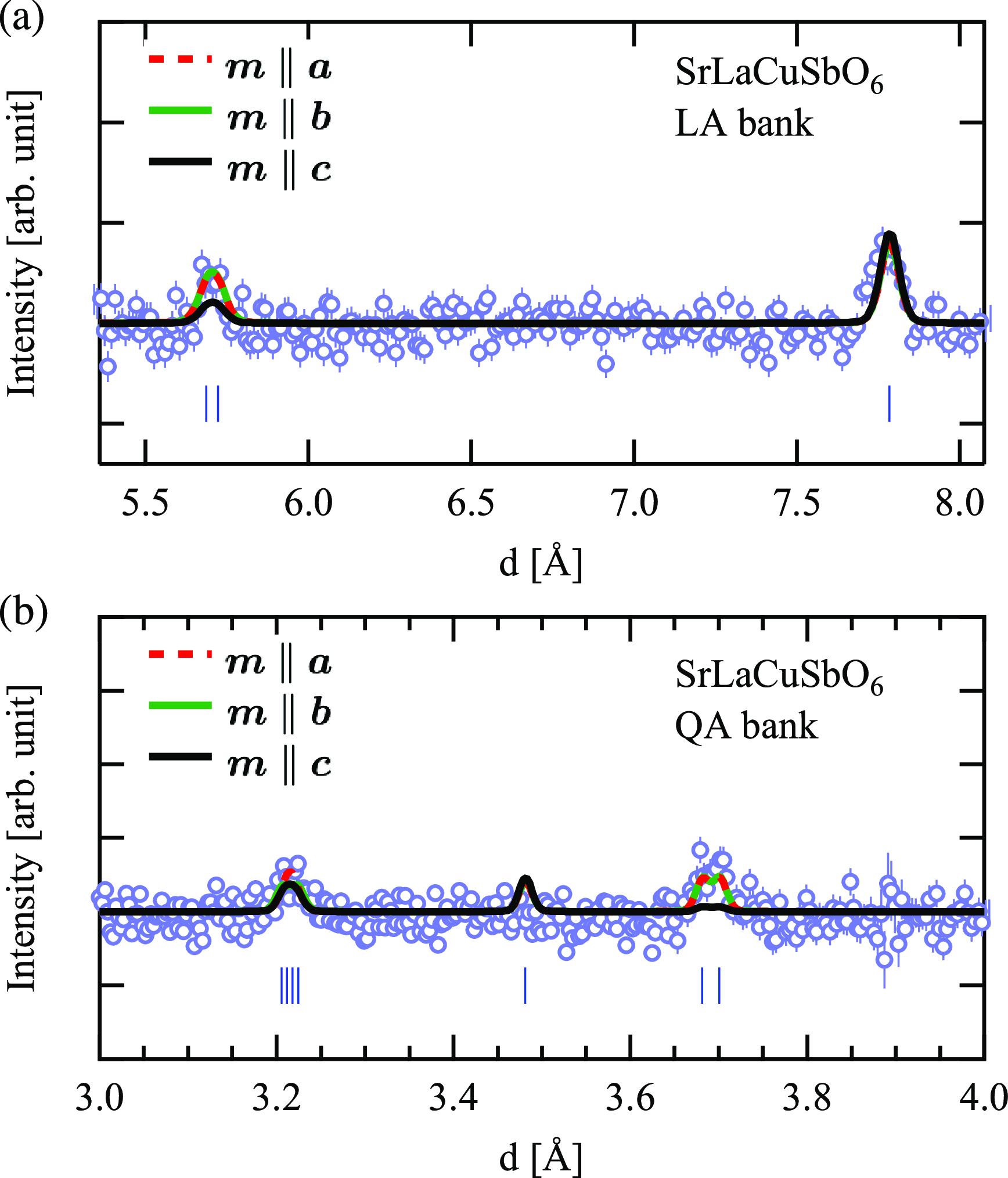}
\end{center}
\vspace{-15pt}\caption{NPD spectra collected from the (a) LA and (b) QA banks at 3.5\,K for SLCSO, where the diffraction spectrum at 20\,K was subtracted as the background. The red, green, and black lines are patterns calculated in accordance with the $\Gamma_2$-based model described in the text. Vertical bars are expected reflections.}
\label{npd_mag_slcs}
\end{figure}

Figure~\ref{npd_mag_slcs} shows NPD spectra of SLCSO collected frin the LA and QA banks at 3.5 K, where the diffraction spectrum at 20\,K (${>}\,T_{\rm N}\,{=}\,13.6$\,K) has been subtracted so that only magnetic Bragg peaks are extracted. All the magnetic peaks can be indexed by the propagation vector ${\bm k}\,{=}\,(1/2, 1/2, 0)$ on the body-centered structure. Therefore, the magnetic structure in the two-dimensional (2D) layer of SLCSO is determined to be NAF, as observed in Sr$_2$CuTeO$_6$~\cite{Koga2016}. The absence of superlattice magnetic reflection for the doubling of the lattice constant $c$ is consistent with good two-dimensionality.

\begin{figure}[]
\begin{center}
\includegraphics[width=1.0\linewidth]{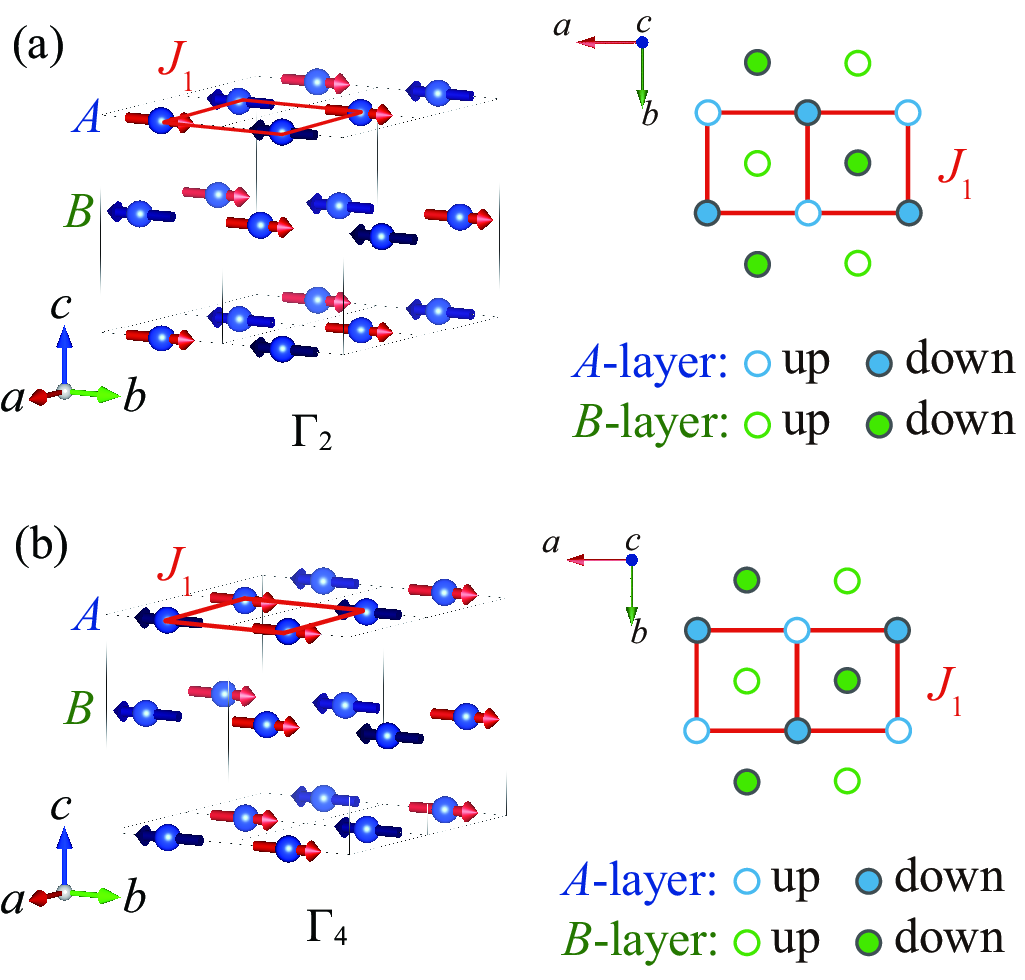}
\end{center}
\vspace{-15pt}\caption{Possible magnetic structures of (a) $\Gamma_2$ and (b) $\Gamma_4$ models for SLCSO obtained by the irreducible representation analysis of the $P2_1/n$ crystal structure.}
\label{model_slcs}
\end{figure}

To determine the magnetic structure in the ordered state of SLCSO, we performed the irreducible representation analysis using the SARAh program~\cite{WILLS2000} for the space group $P2_1/n$ with ${\bm k}\,{=}\,(1/2, 1/2, 0)$. In SLCSO, Cu occupies the Wyckoff $2c$ position, which has two identical sites of $(x, y, z)\,=\,(1/2, 0, 1/2)$ and $(0,1/2,0)$. Because the decomposition of the magnetic representation of the Cu site is expressed as $\Gamma_{\rm mag}=3\displaystyle\Gamma_2^1+3\displaystyle\Gamma_4^1$ in Kovalev's notation, there are two possible models corresponding to $\Gamma_2$ and $\Gamma_4$, as shown in Fig.~\ref{model_slcs}. In the $\Gamma_2$ model, if one sublattice magnetic moment is expressed as $(m_x, m_y, m_z)$, where the $x$-, $y$-, and $z$-axes are chosen to be parallel to the lattice vectors $\bm a$, $\bm b$, and $\bm c$, respectively, the other is expressed as $(m_x, -m_y, m_z)$. On the other hand, in the $\Gamma_4$ model, if one sublattice magnetic moment is expressed by the basis vector $(m_x, m_y, m_z)$, the other is defined by $(-m_x, m_y, -m_z)$. However, from ``powder-averaged" neutron diffraction data, it is difficult to distinguish the $\Gamma_2$ and $\Gamma_4$ models. Thus, to estimate the size of the magnetic moment $m$, we examined the three $\Gamma_2$-based magnetic models, in which the magnetic moments are aligned parallel to the $a$-, $b$-, and $c$-axes. As shown in Fig.~\ref{npd_mag_slcs}, the Rietveld analysis of the magnetic structure reveals that the experimental diffraction patterns are well reproduced by the models with $\bm m\,{\parallel}\,\bm a$ and $\bm m\,{\parallel}\,\bm b$ rather than $\bm m\,{\parallel}\,\bm c$, although the former two models are indistinguishable with the present data. The $R$-factors and magnetic moments for these models are listed in Table~\ref{r_slcs_mag}. From these results, it is concluded that the ordered magnetic moments lie in the $ab$-plane in the magnetically ordered state of SLCSO. The refined Cu$^{2+}$ total magnetic moment, that is, the average of those for the two $\Gamma_2$ models with $\bm m\parallel \bm a$ and $\bm m\parallel \bm b$, is $0.39(3)\,\mu_{\rm B}$, which is considerably smaller than $m\,{=}\,0.69(6)\,\mu_{\rm B}$ observed at 1.6\,K for Sr$_2$CuTeO$_6$~\cite{Koga2016}. We also estimated the size of the magnetic moment, assuming the $\Gamma_4$ model. However, the obtained size of the magnetic moment was the same as that obtained by assuming the $\Gamma_2$ model within error range. It will be possible to determine uniquely the spin structure of SLCSO, if a single crystal is available. However, a high-resolution diffractometer is necessary because the crystal lattice of SLCSO is nearly tetragonal and the difference between the diffraction spectra for the $\Gamma_2$ and $\Gamma_4$ models will be very small.
\begin{table}[]
\caption{$R$-factors and magnetic moments obtained by magnetic structure refinement based on the three $\Gamma_2$ models for SrLaCuSbO$_6$.}
\label{r_slcs_mag}
\begin{ruledtabular}
\begin{tabular}{llllllllll}
&\multicolumn{3}{c}{LA bank}&&\multicolumn{3}{c}{QA bank}&&\\\cline{2-4}\cline{6-8}
Model&$R_{\rm wp}(\%)$ & $R_{\rm e}(\%)$ & $\chi^2$ && $R_{\rm wp}(\%)$ & $R_{\rm e}(\%)$ & $\chi^2$  && $m$\,($\mu_{\rm B}$)\\\hline
$\bm m\parallel \bm a$ & 74.1 & 65.5 & 1.28 && 84.6 & 71.3 & 1.41 && 0.38(2)\\
$\bm m\parallel \bm b$ & 74.2 & 65.5 & 1.28 && 84.0 & 71.3 & 1.39 && 0.39(2)\\
$\bm m\parallel \bm c$ & 76.0 & 65.5 & 1.35 && 95.0 & 71.3 & 1.78 && 0.29(2)\\
\end{tabular}
\end{ruledtabular}
\end{table}

\subsection{Magnetic structure of SrLaCuNbO$_6$}
\begin{figure}[h]
\begin{center}
\includegraphics[width=1\linewidth]{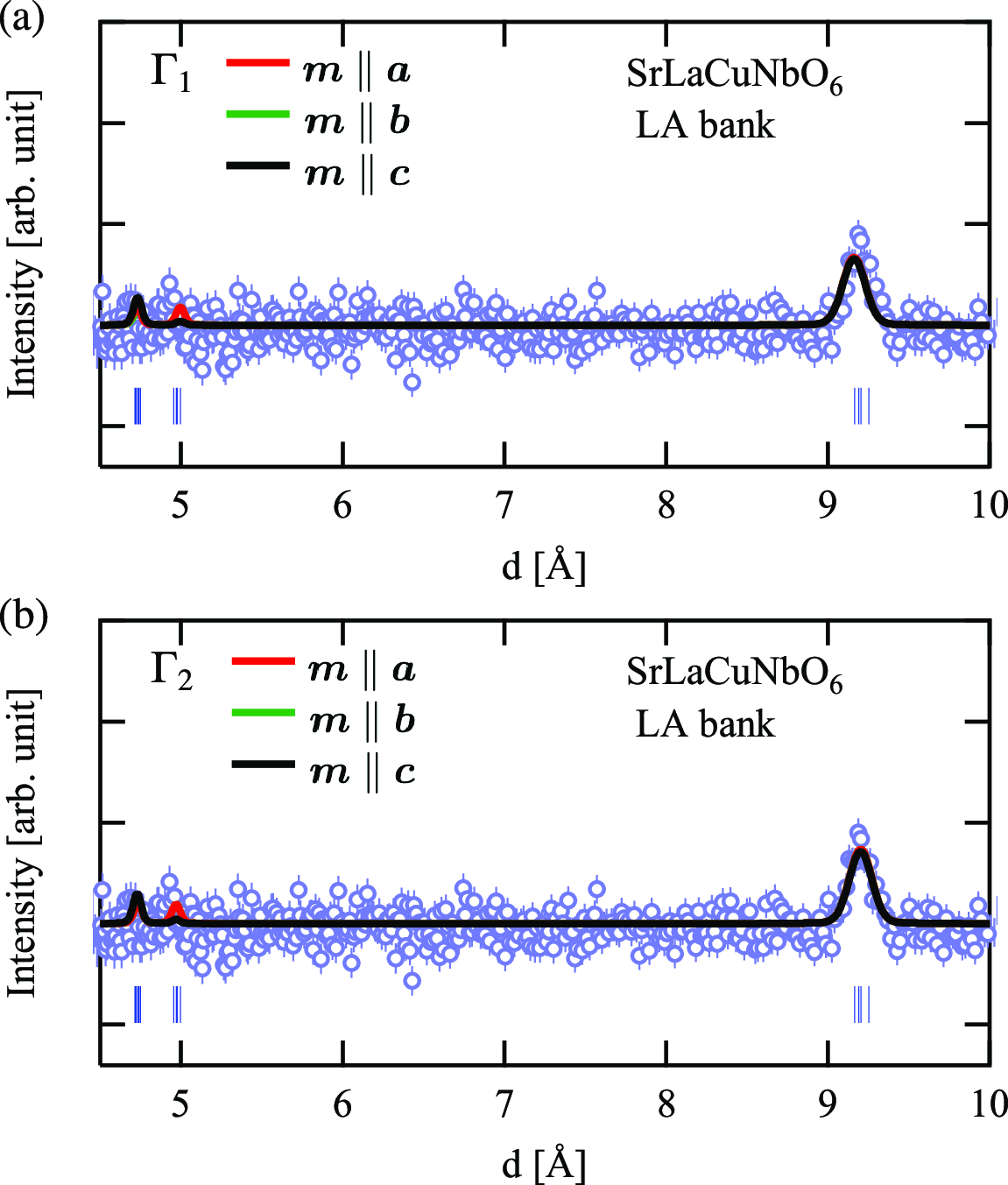}
\end{center}
\vspace{-15pt}\caption{NPD spectra of SLCNO collected from the LA bank at 3.5\,K, where the diffraction spectrum at 20\,K was subtracted as the background. The red, green, and black lines are patterns calculated in accordance with the (a) $\Gamma_1$- and (b) $\Gamma_2$-based models described in the text. Vertical bars are expected reflections.}
\label{npd_mag_slcn}
\end{figure}

Figure \ref{npd_mag_slcn} shows NPD spectra of SLCNO collected from the LA bank at 3.5\,K, where the diffraction spectrum at 20\,K (${>}\,T_{\rm N}\,{=}\,15.7$\,K) has been subtracted so that only magnetic Bragg peaks are extracted. All the magnetic peaks can be indexed by the propagation vector ${\bm k}\,{=}\,({-}1/2, 1/2, 1/2)$ on the face-centered structure. This indicates that the magnetic structure in the 2D layer of SLCNO is CAF, as observed in Sr$_2$CuWO$_6$~\cite{Vasala2014}. Note that there are other possibilities for the propagation vectors, such as ${\bm k}\,{=}\,(1/2, 1/2, -1/2)$, $(1/2, -1/2, 1/2)$, and so on, because the crystal lattice of SLCNO is triclinic. However, it is difficult to distinguish these propagation vectors within experimental error range, because $a\,{\approx}\,b$ and ${\alpha}\,{\approx}\,{\beta}\,{\approx}\,{\gamma}\,{\approx}\,90^{\circ}$, as shown in Table~\ref{slcn_str}.

\begin{figure}[h]
\begin{center}
\includegraphics[width=1\linewidth]{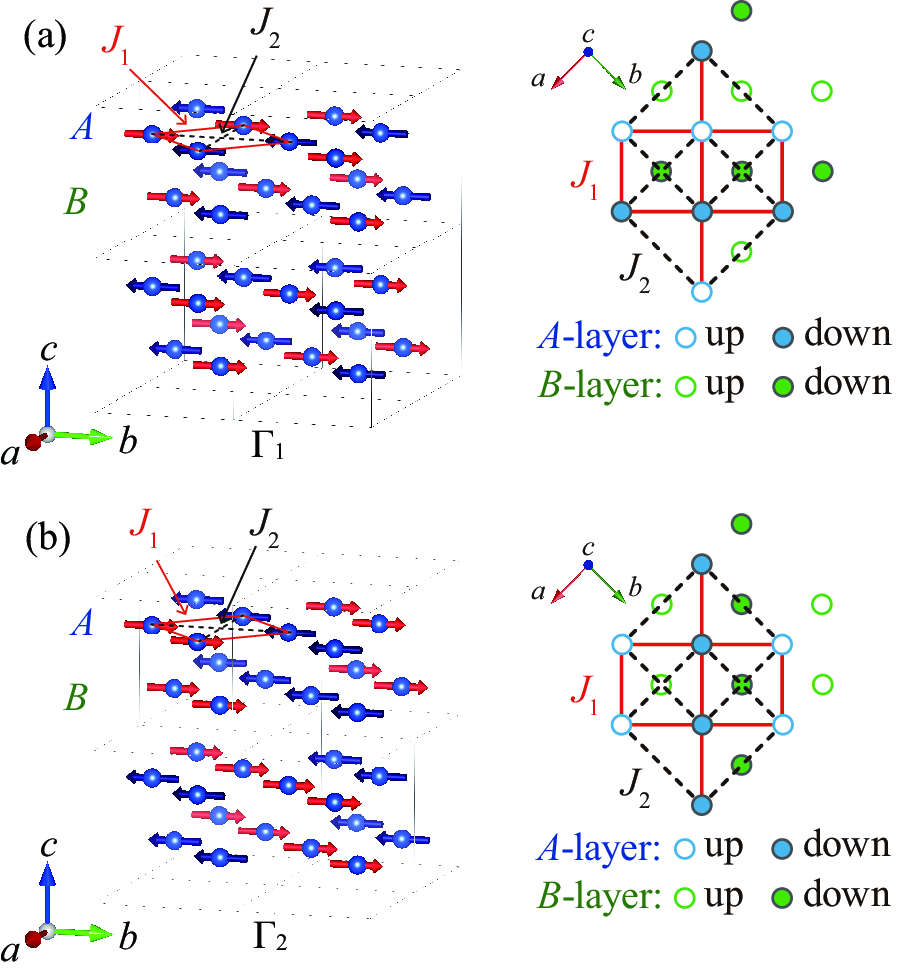}
\end{center}
\vspace{-15pt}\caption{Possible magnetic structures of (a) $\Gamma_1$ and (b) $\Gamma_2$ models for SLCNO obtained by the irreducible representation analysis of the $P{\bar 1}$ crystal structure.}
\label{model_slcn}
\end{figure}

To determine the possible magnetic structures of SLCNO, irreducible representation analysis using the SARAh program \cite{WILLS2000} was performed for the $P{\bar 1}$ crystal structure. In SLCNO, both Cu1 and Cu2  occupy the Wyckoff $2i$ position, which has two identical sites of $(x, y, z)\,=\,(X, Y, Z)$ and $(-X, -Y, -Z)$. The magnetic representation of both the Cu1 and Cu2 sites are described by $\Gamma_{\rm mag}=3\Gamma_1^1+3\Gamma_2^1$ in Kovalev's notation for ${\bm k}\,{=}\,({-}1/2, 1/2, 1/2)$ in the $P{\bar 1}$ crystal structure. This indicates that there are two possible magnetic structures corresponding to $\Gamma_{\rm mag}\,{=}\,3\Gamma_1$ and $3\Gamma_2$, assuming that magnetic moments on the Cu1 and Cu2 sites belong to the same irreducible representation, as shown in Fig.~\ref{model_slcn}.  In the $\Gamma_1$ model, if one sublattice magnetic moment is expressed as $(m_x, m_y, m_z)$, where the $x$-, $y$-, and $z$-axes are chosen to be parallel to the lattice vectors $\bm a$, $\bm b$, and $\bm c$, respectively, the other is expressed as $(m_x, m_y, m_z)$. On the other hand, in the $\Gamma_2$ model, if one sublattice magnetic moment is expressed by the basis vector $(m_x, m_y, m_z)$, the other is given by $(-m_x, -m_y, -m_z)$. Spins align ferromagnetically along the $[1, -1, 0]$ and $[1, 1, 0]$ directions in the $ab$-plane for the $\Gamma_1$ and $\Gamma_2$ models, respectively. These two models are degenerate when the NN interactions $J_1$ and $J_1^{\prime}$ along the $[1, -1, 0]$ and $[1, 1, 0]$ directions, respectively, are the same as in the case of the tetragonal crystal lattice. For SLCNO with a triclinic crystal structure, $J_1\,{\neq}\,J_1^{\prime}$, although they are close to each other; thus, the $\Gamma_1$ and $\Gamma_2$ models will have slightly different ground state energies.

For both the $\Gamma_1$ and $\Gamma_2$ models, a strong diffraction peak is observed only at $d\,{\approx}\,9.2$\,{\AA}. This peak is $(1/2, 1/2, {-}1/2)$ reflection with $d\,{=}\,9.165$\,{\AA} for the $\Gamma_1$ model, while for the $\Gamma_2$ model, it is $({-}1/2, 1/2, 1/2)$ reflection with $d\,{=}\,9.205$\,{\AA}. The $d$-spacing for the $({-}1/2, 1/2, 1/2)$ reflection is slightly larger than that for $(1/2, 1/2, {-}1/2)$ reflection due to the triclinic crystal lattice. The $R$-factors of the refinement and magnetic moments for the $\Gamma_1$ and $\Gamma_2$ models are listed in Table \ref{r_slcn_mag}.We can see that $R$-factor and ${\chi}^2$ for the $\Gamma_2$ model are significantly smaller than those for the $\Gamma_1$ model. Thus, we can deduce that the $\Gamma_2$ model reproduces the experimental diffraction patterns more accurately than the $\Gamma_1$ model. However, it is difficult to determine the orientation of the ordered moment from the present experiment. To evaluate the magnitude of the ordered magnetic moment, we examined three $\Gamma_2$-based models with magnetic moments parallel to the $a$-, $b$-, and $c$-axes, and took the average. The refined Cu$^{2+}$ total magnetic moment was evaluated to be $0.37(1)\,\mu_{\rm B}$, which is approximately two-thirds of $m\,{=}\,0.57(1)\,\mu_{\rm B}$ observed at 3\,K for Sr$_2$CuWO$_6$ \cite{Vasala2014}.

\begin{table}[h]
\caption{$R$-factors and magnetic moments obtained by magnetic structure refinement based on the $\Gamma_1$ and $\Gamma_2$ models for SrLaCuNbO$_6$.}
\label{r_slcn_mag}
\begin{ruledtabular}
\begin{tabular}{llllllll}
&\multicolumn{3}{c}{$\Gamma_1$}&&\multicolumn{3}{c}{$\Gamma_2$}\\\cline{2-4}\cline{6-8}
Model&$R_{\rm wp}(\%)$ & $\chi^2$\footnotemark[1] & $m$($\mu_{\rm B}$) && $R_{\rm wp}(\%)$ & $\chi^2$\footnotemark[1] & $m$\,($\mu_{\rm B}$)\\\hline
$\bm m\parallel \bm a$ & 73.7 & 1.42 & 0.36(2) && 66.3 & 1.15 & 0.37(1)\\
$\bm m\parallel \bm b$ & 73.2 & 1.40 & 0.36(2) && 66.8 & 1.16 & 0.37(1)\\
$\bm m\parallel \bm c$ & 74.3 & 1.44 & 0.34(1) && 68.7 & 1.23 & 0.35(1)\\
\end{tabular}
\end{ruledtabular}
\footnotetext[1]{$R_{\rm e}\,{=}\,61.9\,\%$}
\end{table}

\subsection{NMR measurement}
\begin{figure}[t]
\begin{center}
\includegraphics[width=0.80\linewidth]{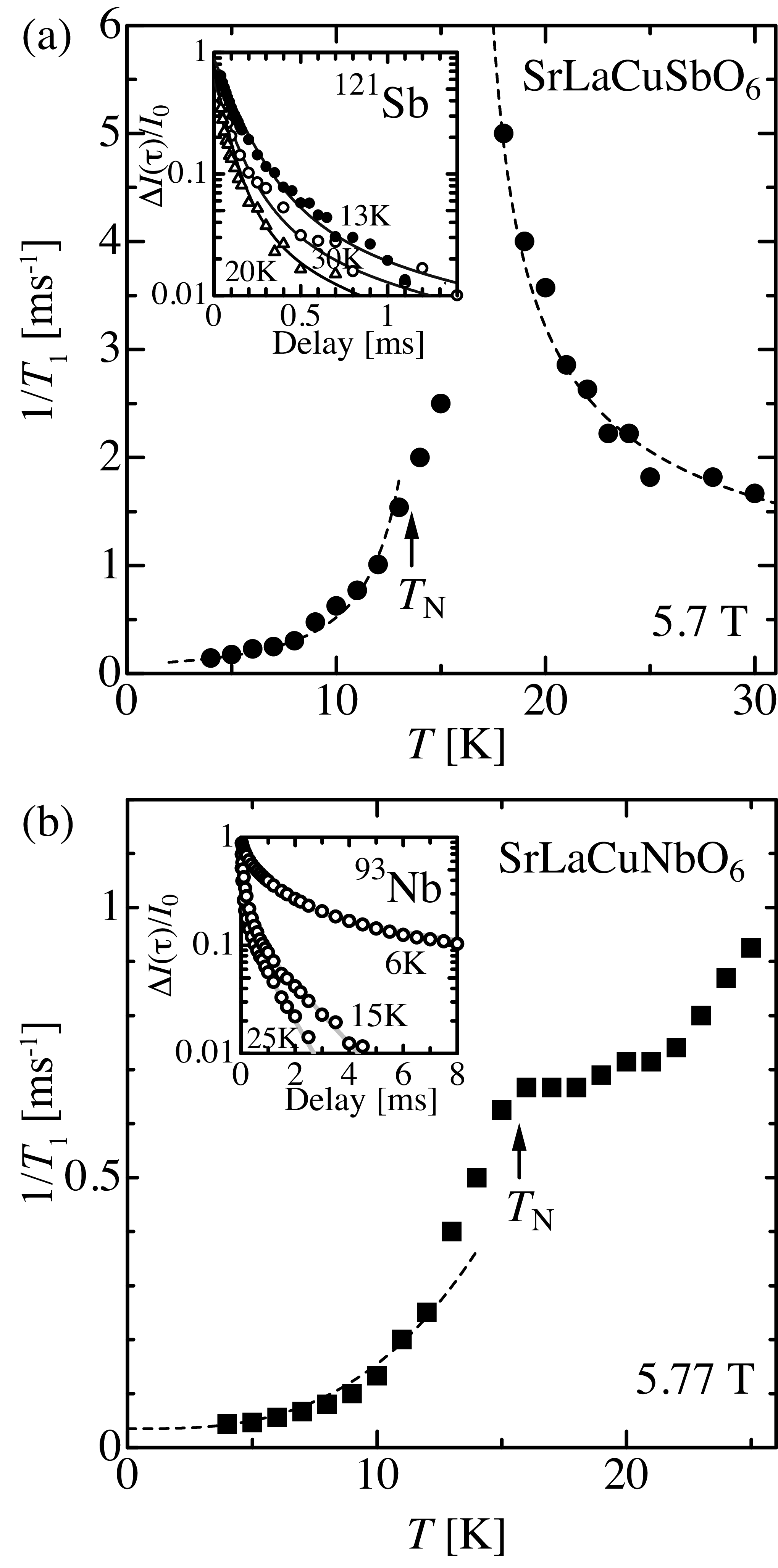}
\end{center}
\vspace{-15pt}\caption{Temperature dependence of $1/T_1$ for (a) $^{121}$Sb nucleus in SLCSO and (b) $^{93}$Nb nucleus in SLCNO.  Dashed curves are a guide to the eyes. Arrows show $T_{\rm N}$ determined by the magnetic susceptibility measurements.  Each inset shows typical recovery curves of nuclear spin magnetization at various temperatures.  Solid curves show theoretical curves for $I\,=\,5/2$ nuclear spins for $^{121}$Sb and for $I\,=\,9/2$ nuclear spins for $^{97}$Nb~\cite{Oosawa2009, Matsui2017}.}
\label{nmr1}
\end{figure}

First, we show the temperature dependence of $1/T_1$ for the two systems in Fig. \ref{nmr1}. Typical recovery curves of nuclear spin magnetization are shown in each inset, with the theoretical curves from which $T_1$ was evaluated. In both systems, $1/T_1$ decreases with decreasing temperature in the paramagnetic state, and after showing divergence for SLCSO or a small hump for SLCNO at $T_{\rm N}$, decreases again to the lowest temperature. The  divergence or hump at $T_{\rm N}$ clearly shows critical slowing down around the second-order phase transition, and hence, ensures that the observed anomalies in the magnetic susceptibility and specific heat are due to the long-range magnetic order. Note that in SLCSO, the temperature region in which $1/T_1$ diverges is slightly higher than $T_{\rm N}$ determined by magnetic susceptibility measurement. This is simply because the critical divergence occurs at the temperature where the mean frequency of the spin fluctuation coincides with the NMR Larmor frequency. This divergence temperature is usually higher than $T_{\rm N}$, at which spin fluctuation freezes completely.

\begin{figure}[h]
\begin{center}
\includegraphics[width=0.80\linewidth]{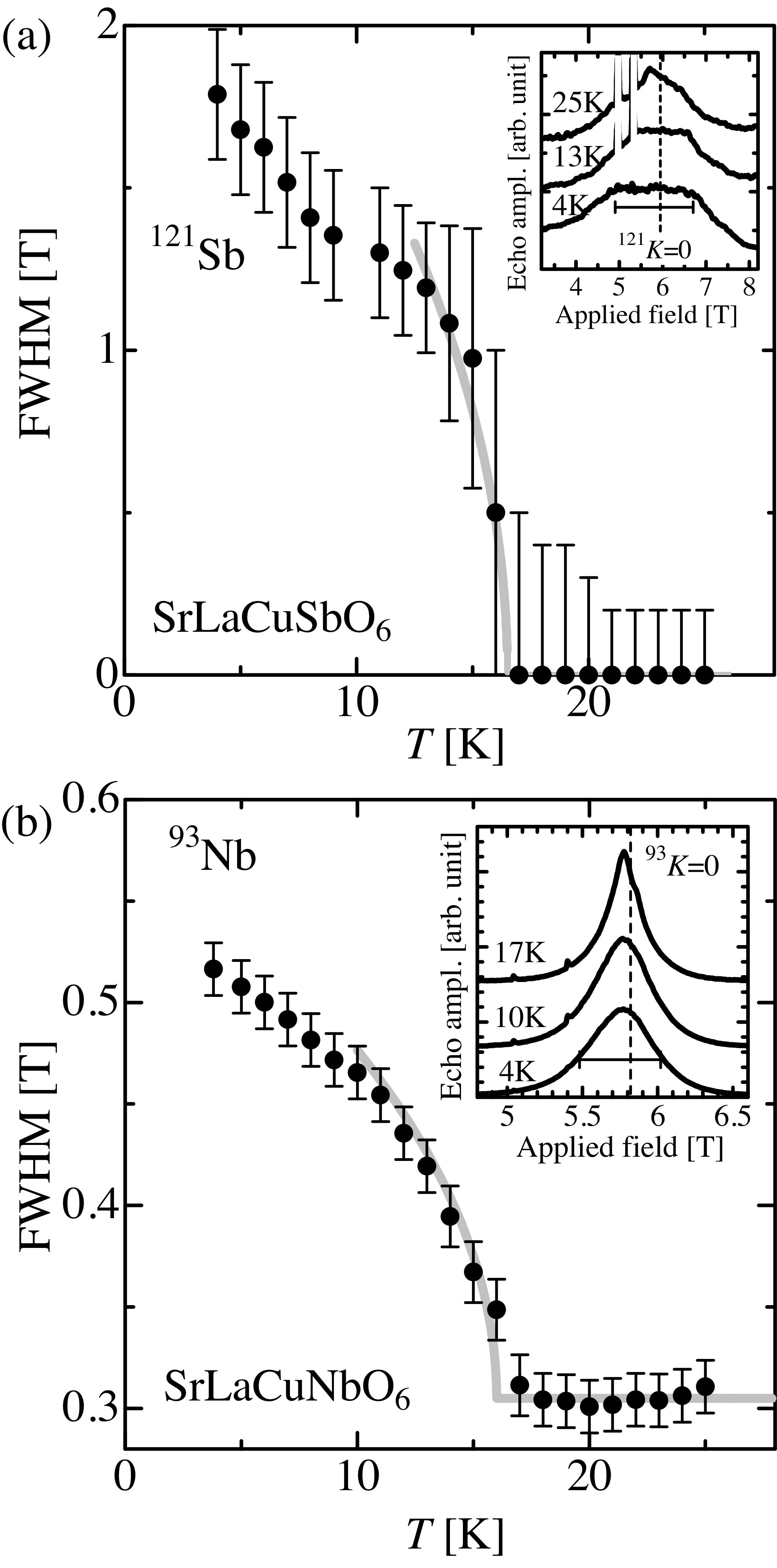}
\end{center}
\vspace{-15pt}\caption{Temperature dependence of static hyperfine field at (a) Sb site for SLCSO and (b) Nb site for SLCNO. Solid curves are a guide to the eyes. Each inset shows typical spectra of the electric quadrupolar powder pattern for the central transition. The NMR zero-shift position is shown by the vertical dashed line. The size of the static hyperfine field in the ordered state shown by the horizontal line with edges is determined as the flat-top width in SLCSO and as the FWHM in SLCNO.} 
\label{nmr2}
\end{figure}

Next, typical spectra at various temperatures below and above $T_{\rm N}$ are shown in the inset of Fig.~\ref{nmr2}. They show appreciable broadening below $T_{\rm N}$, reflecting the long-range ordering of Cu moments. This is particularly noteworthy for SLCSO, showing a flat-top shape in the middle of a powder pattern, which is indicative of 3D antiferromagnetic ordering~\cite{Goto1996}. Although the NMR measurement field is much higher than the spin-flop transition at 1.4\,T, most spins in the powder sample may still be randomly directed randomly to form a 3D powder pattern, because only a limited number of spins whose direction is parallel to the applied field are strongly affected by the spin-flop transition.

The hyperfine field at the NMR site in the ordered state was extracted from the spectra as the width of the flat-top part for SLCSO and as the FWHM for SLCNO~\cite{Matsui2017}. Their temperature dependences are shown in Fig.~\ref{nmr2}, where one can see that the hyperfine field for both systems shows an abrupt increase at $T_{\rm N}$ and continues to increase monotonically to the lowest temperature. The increase in hyperfine field at 4\,K compared with the paramagnetic state markedly differs for the two systems, that is, for SLCSO, it is 1.8\,T, nine times larger than 0.2\,T for SLCNO. Note that if one adopts the FWHM as the width definition for SLCSO, the difference becomes even larger.

\begin{figure}[h]
\begin{center}
\includegraphics[width=0.9\linewidth]{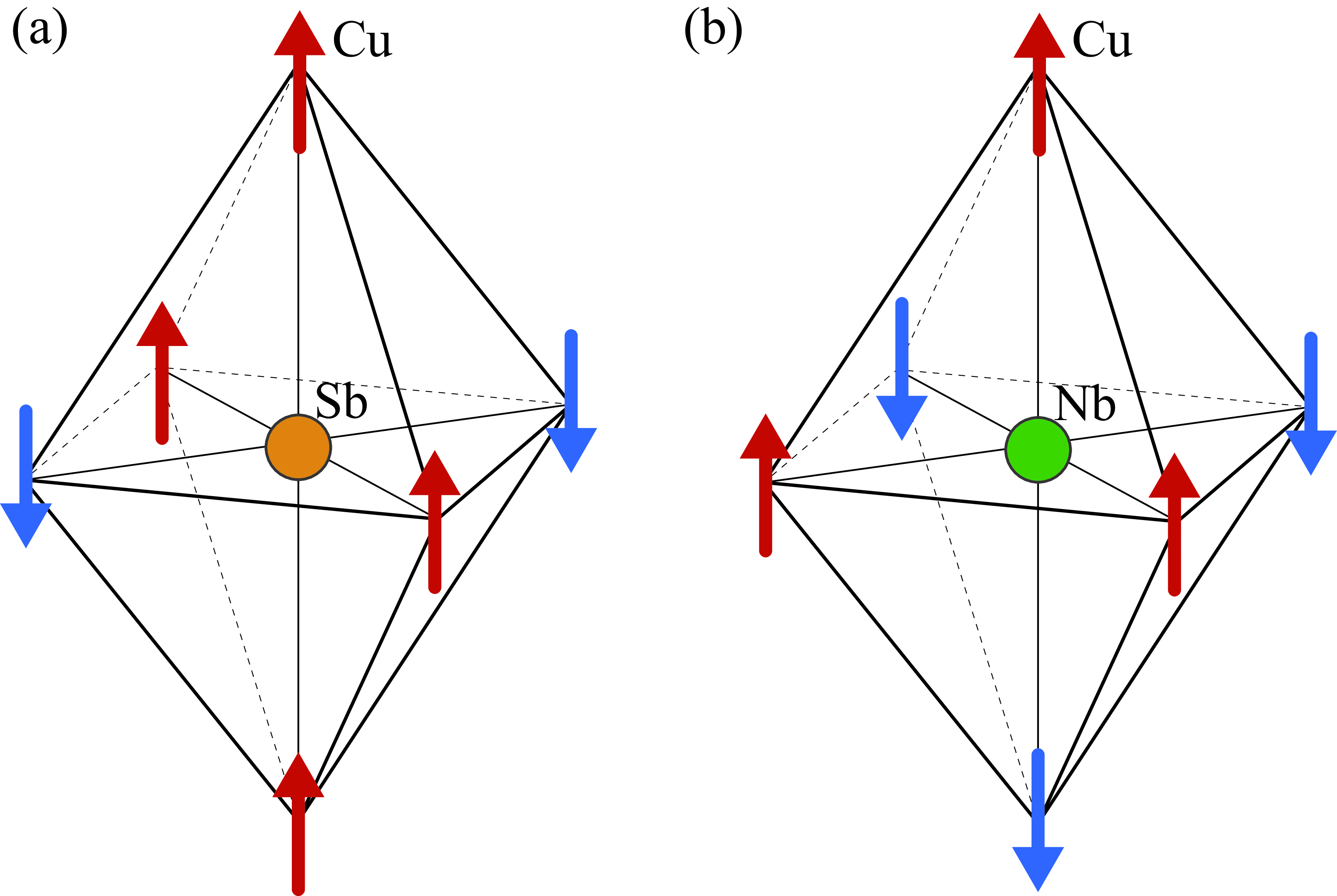}
\end{center}
\vspace{-15pt}\caption{Illustrations of spin configurations and resulting hyperfine fields acting on the NMR sites (a) $^{121}$Sb and (b) $^{93}$Nb in SLCSO and SLCNO, respectively. Two spins facing to each other with NMR sites interposed in between are parallel and antiparallel in SLCSO and SLCNO, respectively. For SLCNO, hyperfine fields from neighboring spins cancel out, while the resulting hyperfine field is finite for SLCSO.} 
\label{canceling}
\end{figure}

Subsequently, we discuss the origin of these differences in the NMR results of the two systems. The two systems have nearly the same $T_{\rm N}$ and are isomorphic, indicating that the strength of hyperfine coupling is also nearly the same. However, there is a large difference both in the size of the static hyperfine field at the lowest temperature and in the strength of the critical divergence at around $T_{\rm N}$. This difference is considered to reflect the different spin structure of each system. That is, when the two spins are located symmetrically around the NMR site, their hyperfine fields cancels out if the two align antiferromagnetically (see Fig.~\ref{canceling}), a phenomenon known as geometrical canceling~\cite{Goto1996}. This cancellation effect also works for dynamically fluctuating spins. If there is a strong antiferromagnetic correlation between the two spins, the resultant hyperfine field may vanish.

Here, we show that the different NMR results observed for the two systems originate from the different spin structures, which are described in the preceding section, via geometrical canceling. That is, the cancelation within the $ab$-plane occurs in both systems, but that along the $c$-axis does not occur for SLCSO, where the two adjacent spins above and below the Sb site align or correlate ferromagnetically, and hence, the hyperfine field from these spins is not canceled out, resulting in a large static hyperfine field in the ordered state and a strong critical divergence in 1/$T_1$ at around $T_{\rm N}$, which are exactly what have been observed. In contrast, the two spins are antiferromagnetically aligned for SLCNO, thus, most of the hyperfine field is canceled out at the Nb site. The small but finite-size hyperfine field observed at the Nb site is simply due to the slight inhomogeneity brought about by the mixing of Sr and La ions~\cite{Goto1996}.

\section{Discussion}\label{discussion}
The present NPD experiments reveal that the magnetic structure in a 2D layer of SLCSO is NAF, which is identical to that of Sr$_2$CuTeO$_6$~\cite{Koga2016}, while the magnetic structure in a 2D layer of SLCNO is CAF, which is identical to that of Sr$_2$CuWO$_6$ \cite{Vasala2014}. These results indicate that the NN interaction $J_1$ and the NNN interaction $J_2$ are dominant in SLCSO and SLCNO, respectively. This difference can be explained by the super-exchange interactions according to the Kanamori theory \cite{KANAMORI1959} as an analog to the cases of  Sr$_2$CuTeO$_6$ and Sr$_2$CuWO$_6$. In SLCSO and SLCNO, the Cu$^{2+}$$-$\,O$^{2-}$$-$\,O$^{2-}$$-$\,Cu$^{2+}$ path is one of the dominant paths between the NN Cu ions (see Figs.~\ref{structure}\,(c) and (d)) and is considered to be antiferromagnetic, as observed in many magnetic materials. The super-exchange interaction via this path gives an antiferromagnetic contribution to the $J_1$ interaction. The other dominant path is Cu$^{2+}$$-$\,O$^{2-}$$-$\,M$^{5+}$$-$\,O$^{2-}$$-$\,Cu$^{2+}$, which is composed of corner-sharing CuO$_6$ and $M$O$_6$ octahedra with $M\,{=}\,$Nb and Sb. The difference in the electronic structure of the nonmagnetic pentavalent Sb$^{5+}$ and Nb$^{5+}$ ions is responsible for the difference in the magnitude of $J_1$ and $J_2$. In SLCSO, the Sb$^{5+}$ outermost occupied 4$d$ orbitals should be core-like below the valence 3$d$ orbital of Cu and the 2$p$ orbital of O as in the case of hexavalent Te$^{6+}$ ions in Sr$_2$CuTeO$_6$ \cite{Xu2017,Xu2018,Babkevich2016}. The orbital hybridization between the 4$d$ orbital of Sb$^{5+}$, the 3$d$ orbital of Cu$^{2+}$, and the 2$p$ orbital of O$^{2-}$ should be small. Therefore, the contribution of the Cu$^{2+}$$-$\,O$^{2-}$$-$\,Sb$^{5+}$$-$\,O$^{2-}$$-$\,Cu$^{2+}$ super-exchange path to the $J_1$ and $J_2$ interactions is expected to be negligible. On the other hand, in SLCNO, since the 5$d$ orbital of Nb$^{5+}$ is unoccupied, the 5$d$ orbital can hybridize with the Cu$^{2+}$ 3$d$ and O$^{2-}$ 2$p$ orbitals in the valence band as in the case of hexavalent W$^{6+}$ ions in Sr$_2$CuWO$_6$ \cite{Xu2017,Xu2018}. This hybridization allows the Cu$^{2+}$$-$\,O$^{2-}$$-$\,Nb$^{5+}$$-$\,O$^{2-}$$-$\,Cu$^{2+}$ super-exchange path, leading to the prominent antiferromagnetic contribution to the $J_2$ interaction. For these reasons, we can deduce that in SLCSO, the NN interaction $J_1$ is strongly antiferromagnetic and the NNN interaction $J_2$ is negligible, while in SLCNO, $J_2$ is stronger than $J_1$, but they are of the same order of magnitude. 

Next, we estimate the exchange constants of SLCSO from the magnetic susceptibility data using the [5, 5] Pad\'{e} approximation combined with the result of the high-temperature series expansion (HTSE) for the magnetic susceptibility of $S\,{=}\,1/2$ $J_1{-}J_2$ SLHAF model up to the tenth order of $J/k_{\rm B}T$ \cite{Rosner2003}. The best fit between 25 and 300 K under the condition of $|J_1|\,{>}\,|J_2|$ is obtained with $J_1/k_{\rm B}\,{=}\,74.7(1)$\,K and $J_2/k_{\rm B}\,{=}\,1.62(4)$\,K using an average $g$-factor of $g_{\rm avg}\,{=}\,2.197$ determined from the ESR measurement. The magnetic susceptibility calculated with these parameters is shown by a blue solid line in Fig.~\ref{mtmh}\,(a). 
The calculated magnetic susceptibility fit well the experimental result, although slight disagreement is observed around the broad anomaly at $T_{\rm max}\,{=}\,71$\,K.

Magnetic specific heat $C_{\rm mag}$ exhibits a rounded maximum at $T^{*}_{\rm max}\,{=}\,44$\,K owing to the short range spin correlation, as shown in Fig.~\ref{hc2}\,(a). The value of the $C_{\rm mag}$ at $T^{*}_{\rm max}$ is $C_{\rm mag}^{\rm max}\,{=}\,0.47\,R$, which coincides with that for an $S\,{=}\,1/2$ SLHAF with the NN interaction~\cite{Bernu2001}. Using a relation $T^{*}_{\rm max}\,{=}\,0.582\,J/k_{\rm B}$ for the $S\,{=}\,1/2$ SLHAF with the NN interaction~\cite{Bernu2001}, we obtain $J/k_{\rm B},{=}\,76$\,K, which is very close to $J_1/k_{\rm B}\,{=}\,74.7$\,K obtained from the analysis of the magnetic susceptibility based on the $S\,{=}\,1/2$ $J_1{-}J_2$ SLHAF model. From these specific heat results, we can deduce that SLCSO is described as a uniform $S\,{=}\,1/2$ $J_1{-}J_2$ SLHAF with dominant NN interaction, and that the exchange constants $J_1/k_{\rm B}\,{=}\,74.7$\,K and $J_2/k_{\rm B}\,{=}\,1.62$\,K obtained from the magnetic susceptibility data are reasonable.

Although the crystal lattice of SLCNO is actually triclinic, it is nearly tetragonal. Thus, we can expect that SLCNO is described as a uniform $S\,{=}\,1/2$ $J_1{-}J_2$ SLHAF as SLCSO. For this reason, we analyze the exchange constants in SLCNO on the basis of $S\,{=}\,1/2$ $J_1{-}J_2$ SLHAF model, assuming that $J_2\,{>}\,J_1$ in contrast to the case for SLCSO with $J_1\,{\gg}\,J_2$. 
The exchange constants $J_1$ and $J_2$ of SLCNO were estimated by the [5, 5] Pad\'{e} approximation using the result of HTSE for the magnetic susceptibility of the $S\,{=}\,1/2$ $J_1{-}J_2$ SLHAF up to the tenth order of $J_n/k_{\rm B}T\ (n\,{=}\, 1$ and 2)~\cite{Rosner2003}. The best fit between 25 and 300 K under the condition of $J_2\,{>}\,J_1$ is obtained with $J_1/k_{\rm B}\,{=}\,18.5(2)$ and $J_2/k_{\rm B}\,{=}\,44.32(5)$\,K using $g_{\rm avg}\,{=}\,2.153$ evaluated from the ESR measurement, which is shown by the green solid line in Fig.~\ref{mtmh}\,(a). The calculated magnetic susceptibility is in good agreement with the experimental data. Thus, it is concluded that SLCNO is described as an $S\,{=}\,1/2$ $J_1{-}J_2$ SLHAF with $J_1/J_2\,{=}\,0.42$.

From the Rietveld analysis, the magnitudes of the ordered moments of SLCSO and SLCNO were found to be $m\,{=}\,0.39(3)\,\mu_{\rm B}$ and $0.37(1)\,\mu_{\rm B}$ at 3.5\,K, respectively. These values are significantly smaller than those of their counterparts Sr$_2$CuTeO$_6$ and Sr$_2$CuWO$_6$, in which $m\,{=}\,0.69(6)\,\mu_{\rm B}$ at\,1.5\,K~\cite{Koga2016} and $0.57(1)\,\mu_{\rm B}$ at 3\,K, respectively~\cite{Vasala2014}. The linear spin wave theory (LSWT)~\cite{Majumdar2010} does not reproduce well the magnitudes of the observed magnetic moments of SLCSO and SLCNO. 
The LSWT calculation for the uniform $S\,{=}\,1/2$ SLHAF gives $m_{\rm calc}\,{=}\,0.303\,g_{\rm avg}\,{=}\,0.666\,\mu_{\rm B}$~\cite{Kubo1952}. The calculated ordered moment is 1.7 times larger than the experimental value for SLCSO. 
For SLCNO, the LSWT calculation for the $S\,{=}\,1/2$ $J_1{-}J_2$ SLHAF with $J_1/k_{\rm B}\,{=}\,18.5$\,K and $J_2/k_{\rm B}\,{=}\,44.3$\,K gives $m_{\rm calc}\,{=}\,0.65\,\mu_{\rm B}$, which is more than twice the observed value. 
We infer that the small observed ordered magnetic moment is ascribed to the bond randomness effect caused by the site disorder of Sr$^{2+}$ and La$^{3+}$ ions. Some theories show that when frustration is weak, the bond randomness does not destroy the magnetic ordering, while it reduces the magnitude of the ordered moment~\cite{Laflorencie2006,Uematsu2018}. It is considered that the site disorder of Sr$^{2+}$ and La$^{3+}$ ions in SLCSO and SLCNO disturbs the local crystal structure and induces weak randomness in the exchange interactions $J_1$ and $J_2$. It is also notable that in Sr$_2$CuTe$_{1{-}x}$W$_x$O$_6$, a small amount of substitution of nonmagnetic hexavalent ion W$^{6+}$ with the different filled outermost orbital from Te$^{6+}$ modifies local exchange interactions and causes drastic reduction of ordering temperature, which leads to QDGS~\cite{Hong2021,Mustonen2018_1,Mustonen2018_2,Watanabe2018,Yoon2021}. The effect of the bond randomness has yet been fully explained.

\section{Conclusion}
We have presented the results of NPD, ESR, NMR, magnetization, and specific heat measurements on SLCSO and SLCNO powders. Magnetization and specific heat measurements show that the quasi-two-dimensional $S\,{=}\,1/2$ $J_1{-}J_2$ SLHAFs SLCSO and SLCNO undergo three-dimensional magnetic ordering at $T_{\rm N}\,{=}\,13.6$ and 15.7\,K, respectively. From NPD measurements, the magnetic structures in 2D layers of SLCSO and SLCNO were determined to be of the N\'{e}el antiferromagnetic (NAF) type characterized by the propagation vector ${\bm k}\,{=}\,(1/2, 1/2, 0)$ on the body-centered structure and of the columnar antiferromagnetic (CAF) type described by the propagation vector ${\bm k}\,{=}\,({-} 1/2, 1/2, 1/2)$ on the face-centered structure, respectively. These magnetic structures were consistent with the results of $^{121}$Sb and $^{93}$Nb NMR measurements. 
The exchange parameters of SLCSO were estimated to be $J_1/k_{\rm B}\,{=}\,74.7$\,K and $J_2/k_{\rm B}\,{=}\,1.62$\,K, which are in the parameter range for the NAF order. On the other hand, the exchange parameters of SLCNO were estimated to be $J_1/k_{\rm B}\,{=}\,18.5$\, K and $J_2/k_{\rm B}\,{=}\,44.3$\,K, which are in the parameter range for the CAF order. The difference in the relative size of $J_2$ to $J_1$ between these two compounds is attributed to whether the electronic configuration of the nonmagnetic pentavalent ions of Nb$^{5+}$ and Sb$^{5+}$ is $d^0$ or $d^{10}$. The magnitudes of the ordered moments of SLCSO and SLCNO were evaluated to be $m\,{=}\,0.39(3)\,\mu_{\rm B}$ and $0.37(1)\,\mu_{\rm B}$ at 3.5\,K, respectively, both of which are significantly smaller than those observed in the related systems Sr$_2$CuTeO$_6$ and Sr$_2$CuWO$_6$, and those calculated by the linear spin wave theory. We infer that the small ordered magnetic moments observed in SLCSO and SLCNO are ascribed to the effect of bond randomness arising from the disorder of Sr$^{2+}$ and La$^{3+}$ ions. SLCSO and SLCNO are thus magnetically described as $S\,{=}\,1/2$ $J_1{-}J_2$ SLHAFs with weak bond randomness.

\section*{ACKNOWLEDGMENTS}

This work was supported by Grants-in-Aid for JSPS Fellows (No.~20J12289) and Scientific Research (A) (No.~17H01142) and (C) (No.~19K03711).

\end{document}